\documentclass[twocolumn]{article}

\pdfoutput=1

\usepackage[utf8]{inputenc}

\usepackage{amsmath} 
\usepackage{amsfonts}
\usepackage{amssymb}

\usepackage[hyphens]{url}

\usepackage{empheq}
\usepackage[export]{adjustbox}

\usepackage{booktabs}
\usepackage{array}
\usepackage{subcaption}

\usepackage{enumitem}
\setlist[itemize]{noitemsep, topsep=0pt}

\title{Epidemiological modeling of the 2005 French riots:\\ a spreading wave and the role of contagion}
\author{Laurent Bonnasse-Gahot\textsuperscript{1},
Henri Berestycki\textsuperscript{1} ,
Marie-Aude Depuiset\textsuperscript{2,3,4},
Mirta B. Gordon\textsuperscript{4},\\
Sebastian Roch\'e\textsuperscript{2},
Nancy Rodriguez\textsuperscript{5},
Jean-Pierre Nadal\textsuperscript{1,6,*}}

\date{{\small
\noindent\textsuperscript{1}\'Ecole des Hautes \'Etudes en Sciences Sociales, PSL Research University, CNRS, Centre d'Analyse et de Math\'ematique Sociales, Paris, France\\
\textsuperscript{2}Univ. Grenoble Alpes, Institut d'Etudes Politiques de Grenoble, CNRS, PACTE, Grenoble, France\\
\textsuperscript{3}Universit\'e de Lille, CNRS, UMR 8019 - CLERS\'E - Centre Lillois d'\'Etudes et de Recherches sociologiques et \'Economiques, Lille, France\\
\textsuperscript{4}Univ. Grenoble Alpes, CNRS, LIG, Grenoble France\\
\textsuperscript{5}The University of North Carolina at Chapel Hill, Department of Mathematics, Chapel Hill, USA\\
\textsuperscript{6}Laboratoire de Physique Statistique, \'Ecole Normale Sup\'erieure, PSL Research University; Universit\'e Paris Diderot, Sorbonne Paris-Cit\'e; Sorbonne Universit\'es, UPMC -- Univ. Paris 06; CNRS; Paris, France.\\
*corresponding author. Email: jpnadal@ehess.fr}\\
$\,$\\
\small
Dated: December 15th, 2017\\
 This is an author-produced version of an article accepted for publication in Scientific Reports (\url{http://dx.doi.org/10.1038/s41598-017-18093-4})
}

\begin{document}

\twocolumn[
\begin{@twocolumnfalse}

\maketitle

\begin{abstract}
As a large-scale instance of dramatic collective behavior, the 2005 French riots started in a poor suburb of Paris, then spread in all of France, lasting about three weeks. Remarkably, although there were no displacements of rioters, the riot activity did travel. Access to daily national police data 
has allowed us to explore the dynamics of riot propagation. Here we show that an epidemic-like model, with just a few parameters and a single sociological variable characterizing neighborhood deprivation, accounts quantitatively for the full spatio-temporal dynamics of the riots. This is the first time that such data-driven modeling involving contagion both within and between cities (through geographic proximity or media) at the scale of a country, and on a daily basis, is performed. Moreover, we give a precise mathematical characterization to the expression ``wave of riots'', and provide a visualization of the propagation around Paris, exhibiting the wave in a way not described before. The remarkable agreement between model and data demonstrates that geographic proximity played a major role in the  
propagation, even though information was readily available everywhere through media. Finally, we argue that our approach gives a general framework for the modeling of the dynamics of spontaneous collective uprisings.\\
\end{abstract}
\end{@twocolumnfalse}
]

\bigskip

\section*{Introduction}
Attracting worldwide media attention, France experienced during the Autumn of 2005 the longest and most geographically extended riot of the  contemporary history of Europe\cite{Waddington_etal_2009,Roche_2010b}. Without any political claims nor leadership, localization was mainly limited to the ``banlieues'' (suburbs of large metropolitan cities), where minority groups are largely confined. Contrary to the London ``shopping riots'' of 2011, rioting in France essentially consisted of car destruction and confrontations with the police. The triggering event took place in a deprived municipality at the north-east of Paris: on October 27, 2005, two youths died when intruding into a power substation while trying to escape a police patrol. Inhabitants spontaneously gathered on the streets with anger. Notwithstanding the dramatic nature of these events, 
the access to detailed police data\cite{Cazelles_etal_2007}, together with the extension in time and space -- three weeks, more than 800 municipalities hit across all of France --, provide an exceptional opportunity for studying the dynamics of a large-scale riot episode. The present work aims at analyzing these data through a mathematical model that sheds new light on qualitative features of the riots as instances of collective human behavior\cite{Raafat_etal_2009,Gross_2011}.

Several works\cite{Stark_etal_1974,Granovetter_1978,Burbeck_etal_1978,Pitcher_etal_1978,Myers_2000,Braha_2012,Baudains_etal_2013,Baudains_etal_2013b,Davies_etal_2013,Berestycki_etal_2015} 
have developed mathematical approaches to rioting dynamics, and their sociological implications have also been discussed\cite{Salgado_etal_2016}. 
The 1978 article of Burbeck {\it et al.}~\cite{Burbeck_etal_1978} pioneered quantitative epidemiological modeling to study the dynamics of riots. Very few works followed the same route, but similar ideas have been applied to other social phenomena such as the spreading of ideas or rumors\cite{Dietz_1967}  and the viral propagation of {\em memes} on the Internet\cite{Wang_Wood_2011}. This original epidemiological modeling was however limited to the analysis within single cities, without spatial extension. 
From the analysis of various sources, previous historical and sociological studies have discussed riot contagion from place to place
\cite{Midlarsky_1978,Govea_West_1981,Bohstedt_Williams_1988,Charlesworth_1994,Myers_2010}. 
However few studies aim at quantitatively describing the spatial spread of riots, 
except for two notable exceptions.  
Studies of the 2011 London riots\cite{Baudains_etal_2013b,Davies_etal_2013} describe the displacements of rioters from neighborhoods to neighborhoods. In contradistinction with the London case, media reports and case studies\cite{Cazelles_etal_2007,Mazars_2007} show that the 2005 French rioters remained localized in a particular neighborhood of each municipality. However, the riot itself did travel.
Conceptualizing riots as interdependent events, Myers makes use of   
the {\em event history approach}\cite{Strang_Tuma_1993} to study the US ethnic riots on a period of several years. This analysis exhibits space-time correlations showing that riots diffused from cities to cities\cite{Myers_1997, Myers_2000, Myers_2010}. There, each rioting episode is considered as a single global event (whether the city ``adopts a riot'' or does not), and measures of covariances allow to relate the occurrence of a riot in a city at a given time with the occurrence of riots in other cities at previous times. 
This approach however does not describe the internal dynamics of a riot (its rise and fall within each city), nor the precise timing of the spread from city to city. Of course, going beyond this framework requires much more detailed data. 

Our dataset, at a level of detail hitherto unavailable, allows us to provide the first data-driven modeling of riot contagion from city to city at the level of a whole country, coupled with contagion within each city, and with a time resolution of a day. Our work, of a different nature than that of the econometric one, 
takes its root in the epidemiological approach introduced in the seminal work of Burbeck {\it et al}~\cite{Burbeck_etal_1978}, and is in the spirit of recent continuous spatio-temporal data-driven approaches in social science\cite{Mohler_etal_2011,Brantingham_etal_2012,Gauvin_etal_2013,Davies_etal_2013}. 
Here we extend the notion of epidemiological propagation of riots by including spatial spreading, in a context where there is no displacement of rioters. Remarkably, the high quality of our results is achieved within the sole epidemiological framework, without any explicit modeling of, e.g., the police actions (in contrast with the 2011 London riots modeling\cite{Davies_etal_2013}).  For the first time, the present study provides a  spatio-temporal framework that shows that, following a specific triggering event,  propagation of rioting activity is analogous (but for some specificities) to the continuous propagation of epidemics.

More precisely, we introduce here a compartmental epidemic model of the Susceptible-\-Infected-Re\-covered (SIR) type\cite{Kermack_McKendrick_1927,Diekmann_Heesterbeek_2000,Hethcote_2000}. Infection takes place through contacts within cities as well as through other short- and long-range interactions arising from either interpersonal networks or media coverage\cite{Stark_etal_1974,Myers_2010}. These influence interactions are the key to riots spreading over the discrete set of French municipalities. In particular, diffusion based on geographic proximity played a major role in generating a kind of riot wave around Paris which we exhibit here. This is substantiated by the remarkable agreement between the data and the model at various geographic scales. Indeed, one of our main findings is that less than ten free parameters together with only one sociological variable (the size of the population of poorly educated young males) are enough to accurately describe the complete spatio-temporal dynamics of the riots. 

The  qualitative features taken into account by our model -- the role of a single triggering effect, a ``social tension'' buildup, a somewhat slower and rather smooth relaxation, and local as well as global spreading --, are common to many riots. This suggests that our approach gives a general framework for the modeling of the spatio-temporal dynamics of spontaneous collective uprisings.

\section*{Results} 
\subsection*{The 2005 French riots dataset}
We base our analysis here on the daily crime reports\cite{Cazelles_etal_2007} of all incidents recorded by the French police at the municipalities (corresponding to the French ``communes'') under police authority, which cover municipalities with a population of at least 20,000 inhabitants. Such data, on the detailed time course of riots at the scale of hours or days, and/or involving a large number of cities, are rare. In addition, as an output of a centralized national recording procedure applied in all national police units operating at the local level, the data are homogeneous in nature -- and not subject to the selection or description biases which are frequent with media sources\cite{Earl_etal_2004}. These qualities endow these data with a unique scientific value. We adopt a simple methodology for quantifying the rioting activity: we define as a single event any rioting-like act, as listed in the daily police reports, leaving aside its nature and its apparent intensity. Thus, each one of ``5 burnt cars''
, ``police officers attacked with stones'' or ``stoning of firemen'', is labeled as a single event. We thus get a dataset composed of the number of riot-like events for each municipality, every day from October 26 to December 8, 2005, a period of 44 days which covers the three weeks of riots and extends over two weeks after.  

Figure~\ref{fig:single_site_fits}a (left panel) shows at its top two typical examples of the time course of the number of events for municipalities (see also the plots for the $12$ most active \^Ile-de-France municipalities, Supplementary Fig. S1). A striking observation is that there is a similar up-and-down dynamics at every location, showing no rebound, or, if any, hardly distinguishable from the obvious stochasticity in the data. This pattern is similar to the one observed for the US ethnic riots\cite{Burbeck_etal_1978}. 
In addition, as illustrated on Fig.~\ref{fig:single_site_fits} and Supplementary Fig. S3, we observe the same pattern across different spatial scales (municipalities, {\em d\'epartements}, {\em r\'egions}, all country -- see Materials and Methods for a description of these administrative divisions). Moreover, this pattern shows up clearly despite the difference in amplitudes (see also 
section \emph{Fitting the data: the wave across the whole country}).
%% Fig.~\ref{fig:fit_dpt_summary}a and \ref{fig:fit_dpt_summary}b). 
This multi-scale property suggests an underlying mechanism for which geographical proximity matters. Finally, the rioting activity appears to be on top of a background level: as can be seen on Fig.~\ref{fig:single_site_fits}, the number of events relaxes towards the very same level that it had at the outset of the period. Actually, in the police data, one cannot always discriminate rioting facts from ordinary criminal ones, such as the burning of cars unrelated to collective uprising. For each location, we assume that the stationary background activity corresponds to this ``normal'' criminal activity. 

\subsection*{Modeling framework}
We now introduce our modeling approach. Section Materials and Methods provides the full model and numerical details, as well as various quantitative statistical analyses for the fits that follow. The model features presented below are based on the analysis at the scale of municipalities. However, since aggregated data at the scale of d\'epartements present a pattern similar to the data of  municipalities, we also fit the model at the d\'epartement scale, as if the model assumptions were correct at the scale of each d\'epartement. A ``site'', below, is either a municipality or a d\'epartement depending on the scale considered.\\
As the rioting activities are described by a discrete set of events, we assume an underlying point process\cite{Baddeley_2007} characterized by its mean value. Assuming no coupling between the dynamics of the rioting and criminal activities (see Materials and Methods for a discussion), the expected number of events at each site $k$ ($k=1,...,K$, $K$ being the number of sites), is the sum of the mean (time independent) background activity $\lambda_{b k}$, and of the (time dependent) rioting activity, $\lambda_k(t)$. In fitting the model to the data, we take the background activity $\lambda_{b k}$ as the average number of events at the considered site over the last two weeks of our dataset. Assuming Poisson statistics (which appears to be in good agreement with the data, see Materials and Methods), the means  $\lambda_k(t)$ fully characterize the rioting activities. We make the assumption that this number of events $\lambda_k(t)$ is proportional to the local number of rioters, $I_k(t)$:
\begin{equation}
\lambda_k(t) = \alpha I_k(t).
\end{equation} 
We model the coupled dynamics of the set of $2 \times K$ variables, the numbers  
$I_k(t)$ of rioters ({\it infected} individuals in the terminology of the SIR model) and the numbers $S_k(t)$ of individuals {\it susceptible} to join the riot, by writing an epidemic SIR model\cite{Kermack_McKendrick_1927, Hethcote_2000} in a form suited for the present study, as explained below. 
This gives the coupled dynamics of the $\lambda_k(t)$ and of the associated variables, 
\begin{equation}
\sigma_k(t)\equiv \alpha S_k(t). 
\end{equation}
These dummy variables can be seen as the reservoirs of events (the maximum expected numbers of events that may occur from time $t$ onwards).  
We fit the model to the data by considering a discrete time version of the equations (events are reported on a daily basis), and by optimizing the choice of the model free parameters with a maximum likelihood method. 
The result of the fit is a set of $K$ smooth curves (in time), $\lambda_k(t), k=1,.., K$. For each location $k$, and each time $t$, the corresponding empirical data point has to be seen as a probabilistic realization of the Poisson process whose mean is $\lambda_k(t)$.

Before going into the modeling details and the fits, we now give the main characteristics of the proposed SIR model. We assume homogeneous interactions {\em within} each municipality (a hypothesis justified by the coarse-grained nature of the data, and by the absence of displacements of rioters), and influences between sites. The model thus belongs to the category of metapopulation epidemic models~\cite{Ball_etal_2015}. Motivated by the relative smoothness of the time course of events, we make the strong assumption that, at each site, there is a constant rate at which rioters leave the riot. This parameter aggregates the effects of different factors -- arrests, stringent policing, other sources of deterrence, fear, fatigue, etc. --, none of them being here modeled explicitly. In addition, since there are almost no rebounds of rioting activity, we assume that there is no flux from {\em recovered} (those who left the riot) to {\em susceptible} (and thus we do not have to keep track of the number of recovered 
individuals). 

In the epidemic of an infectious disease, contagion typically occurs by dyadic interactions, so that the probability for a susceptible individual to be infected is proportional to the {\em fraction} of infected individuals -- leading to equations written in terms of the fractions of infected and susceptible individuals. In the present context, contagion results from a bandwagon effect\cite{Schelling_1973,Granovetter_1978,Raafat_etal_2009}. The probability of becoming a rioter is thus a function of the {\em  number} of rioters, hence of the number of events given the above hypothesis. This function is non-linear since, being a probability, it must saturate at some value (at most 1) for large rioting activities.

\subsection*{Single site epidemic modeling}
As a first step, following Burbeck {\it et al.}, we ignore interactions between sites, and thus specify the SIR model for each site {\em separately}. We consider here one single site (and omit the site index $k$ in the equations). Before a triggering event occurs at some time $t_0$, there is a certain number $S_0 > 0$ of susceptible individuals but no rioters. At $t_0$ there is an exogenous shock leading to a sudden increase in the $I$ population, hence in $\lambda$, yielding an initial condition $\lambda(t_0)=A >0$. From then on, the rioting activity at a single (isolated) site evolves according to:
\begin{subequations}
	\begin{empheq}[left=\empheqlbrace]{align}
	\frac{d \lambda(t)}{dt} &= - \omega \, \lambda(t) + \beta \, \sigma(t) \, \lambda(t), \\
	\frac{d \sigma(t)}{dt} &= - \beta \, \sigma(t) \, \lambda(t), 
	\end{empheq}
\end{subequations}
where $\beta$ is a susceptibility parameter. Here we work within a linear approximation of the probability to become infected, which appears to provide good results for the single site modeling The condition for the riot to start after the shock is that the reproduction number\cite{Diekmann_Heesterbeek_2000} $R_0 = \beta \sigma(t_0)/\omega$ is greater than $1$. In such a case, from $t=t_0$ onward, the number of infected individuals increases, passes through a maximum and relaxes back towards zero. 

We obtain the initial condition $\sigma_0=\sigma(t_0)=\alpha S_0$ from the fitting procedure. Thus for each site, we are left with five free parameters to fit in order to best approximate the time course of the rioting events: $\omega$, $\beta$, $t_0$, $A$ and $\sigma_0$. 

\begin{figure}
	\begin{minipage}{\linewidth}
		\centering
		\includegraphics[width=\linewidth]{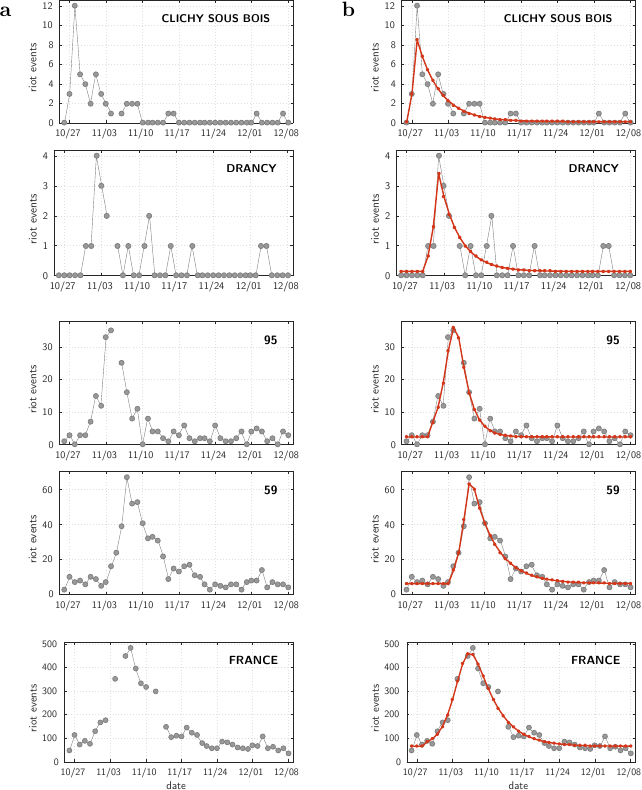}
		\caption{Data and single site fits at different scales. 
			(a) Raw data (gray dots), provided alone for a clearer and unbiased view. (b) Same data
			along with the calibrated model (red curve). Top: municipalities; Middle: d\'epartements; Bottom: all of France (see Materials and Methods for a description of these administrative divisions). Here and in all the other figures involving time, the thin dotted lines divide the time axis into one week periods, starting from the date of the shock, October 27, 2005.}
		\label{fig:single_site_fits}
	\end{minipage}
\end{figure}

By showing examples at different scales, Fig.~\ref{fig:single_site_fits} (b, red curves) illustrate the remarkable quality of the resulting fits (see also Supplementary Fig. S1). The obvious limitation is that fitting all the $853$ municipalities present in the dataset amounts to  determining $853 \times 5 = 4265$ free parameters. The fit is very good but meaningless (overfitting) for sites with only one or two events. In addition, these single site fits cannot explain why the riot started on some particular date at each location. Fitting the single site model requires one to assume that there is one exogenous specific shock at a specific time at each location, whereas the triggering of the local riot actually results from the riot events that occurred before elsewhere. Nevertheless, we see that everywhere the patterns are compatible with an epidemic dynamics and that through the use of the model it is possible to fill in missing data and to smooth the data (filtering out the noise). As a result of this 
filtering, the global pattern of propagation becomes more apparent. Indeed, looking at the Paris area, one observes a kind of wave starting at Clichy-sous-Bois municipality, diffusing to nearby locations, spreading around Paris, and eventually dying out in the more wealthy south-west areas (see Supplementary Video 1). 

\subsection*{Modeling the riot wave}
We now take into account the interactions between sites, specifying the global metapopulation SIR model. 
Among the $K$ sites under consideration, only one site $k_0$, the municipality of Clichy-sous-Bois (d\'epartement 93 when working at d\'epartement scale), undergoes a shock at a time $t_0$, October 27, 2005. To avoid a number of parameters which would scale with the number of sites, we choose here all free parameters to be site-independent (in Materials and Methods we give a more general presentation of the model). The resulting system of $2\times K$ coupled equations writes as follows: for $t>t_0$, for $k=1,...,K$, 
\begin{subequations}
	\begin{empheq} 
	[left=\empheqlbrace]{align} 
	\frac{d \lambda_k(t)}{dt} &=\; - \omega \, \lambda_k(t) + \sigma_k(t) \, \Psi(\Lambda_k(t)), \\
	\frac{d \sigma_k(t)}{dt} &= \;- \sigma_k(t) \, \Psi(\Lambda_k(t)).
	\end{empheq}
\end{subequations}
Here $\omega$ is the site-independent value for the recovering rate. 
For the interaction term we consider that at any site $k$ the probability to join the riot is a function $\Psi$ of a quantity $\Lambda_k(t)$, the global activity as ``seen'' from site $k$. This represents how, on average, susceptible individuals feel concerned by rioting events occurring either locally, in neighboring cities, or anywhere else in France. 
Whatever the means by which the information on the events is received (face-to-face interaction, phone, local or national media -- TV or radio broadcasts, newspapers --, digital media, ...), 
we make the hypothesis that the closer the events (in geographic terms), the stronger their influence. 
We thus write that $\Lambda_k(t)$ is a weighted sum of the rioting activities occurring in all sites, 
\begin{equation}
\Lambda_k(t) = \; \sum_j W_{kj}\, \lambda_j(t),
\end{equation}
where the weights $W_{kj}$ 
depend on the distance between sites $k$ and $j$. 
A simple hypothesis would have been to assume nearest-neighbor contagion. 
We have checked that such scenario fails to reproduce the riots dynamics, which can be easily understood: the riot would not propagate from areas with deprived neighborhoods to other similar urban areas whenever separated by cities without poor neighborhoods. 
We rather consider the weights as given by a decreasing 
function of the distance. We tested several ways of choosing this function 
and obtained the best results for two types of parameterization. One is a power law decay with the distance, motivated by several empirical studies of interactions relying on modern technologies\cite{Liben-Nowell_etal_2005,Lambiotte_etal_2008,Goldenberg_Levy_2009}. The second option is the sum of an exponential decay and of a constant term. Both involve two parameters, a proximity scale $d_0$ and, respectively, the exponent $\delta$ and the strength $\xi$ of the constant term.\\
For the (site independent) function $\Psi(.)$, we consider either its linear approximation, writing
\begin{equation}
\Psi(\Lambda_k(t)) \; = \;  \beta \, \Lambda_k(t) \; = \; \beta \, \sum_j W_{kj}\, \lambda_j(t),
\end{equation}
with the susceptibility $\beta$ as a site-independent free parameter, or various non-linear cases, involving up to four parameters. 

Lastly, we have to make the crucial choice of the initial values $\sigma_{k,0}=\sigma_k(t_0)$, specific to each site. By definition, they must be proportional to the size of the initial susceptible population. We make the hypothesis that the latter scales with the size of a population defined by a sociological specification. Thus we assume 
\begin{equation}
\sigma_{k,0} = \zeta_0 \; N_k,
\end{equation} 
where $\zeta_0$ is a site-independent free parameter, and $N_k$ is the size of a reference population provided 
for each municipality by INSEE, the French national statistics and economic studies institute. The results we present below take as reference the population of males aged between 16 and 24 out-of-school with no diploma. 
We find this population, whose size can be viewed as an index of deprivation, to provide the best results when comparing the model fits done with different reference populations (see Materials and Methods). 
This is in line, not only with the fact that riots started and propagated in poor neighborhoods, but also with the fact that most rioters where males, young, and poorly educated\cite{Cazelles_etal_2007,Waddington_etal_2009,Roche_2010b} -- features common to many urban riots\cite{Gross_2011}. One should note that, once we have chosen this specific reference population -- hence setting the susceptible population in deprived neighborhoods --, the hypotheses on the structure of the interactions implicitly assume interactions between populations with similar socio-economic characteristics. In particular, a distance-independent term in the interaction weights may correspond to proximity primarily perceived in terms of cultural, socio-economic characteristics. Our model thus allows to combine spatial and socio-economic characteristics, which are both known to potentially affect riot contagion\cite{Myers_2000,Braun_Koopmans_2010}.

Finally, for the whole dynamics (with a number of coupled equations ranging from $186$ up to $2560$, depending on the case, see below), in the simplest linear case we are left with only six free parameters: $\omega$, $A$, $\zeta_0$, $d_0$, $\delta$ or $\xi$, and $\beta$. 
In the non-linear case, we have five parameters as for the linear case, $\omega$, $A$, $\zeta_0$, $d_0$, $\delta$ or $\xi$, and, in place of $\beta$, up to four parameters depending on the choice of the function $\Psi$. In the following, we will also allow for specific $\beta$ values at a small number of sites, adding as many parameters.

The above model, in the case of the linear approximation, makes links to the classical spatially continuous, non-local, SIR model\cite{Kendall_1957} (see section \emph{Links to the original spatially continuous SIR model} in Materials and Methods, and Supplementary Videos 3 and 4). In dimension one, when the space is homogeneous,  we know\cite{Kendall_1965} that traveling waves can propagate, quite similar to the way the riot spread around Paris as exhibited in the previous section. The new class of models we have introduced is however somewhat different and more general, and raises several open mathematical questions. The next section shows the wave generated by our global model and the fit to the data.

\subsection*{Fitting the data: the wave around Paris}
We first focus on the contagion around Paris, characterized by a continuous dense urban fabric with deprived neighborhoods. There are $1280$ municipalities in  \^Ile-de-France. Among the ones under police authority (a total of $462$ municipalities, for all of which we have data), $287$ are mentioned for at least one riot-like event. For all the other municipalities, which are under ``gendarmerie'' authority (a military status force with policing duties), we have no data. Since their population size is small, we expect the associated numbers of riot events to be very small if not absent, so that these sites have little influence on the whole dynamics. We choose the free parameters with the maximum likelihood method, making use of the available data, i.e. the $462$ municipalities. However, the model simulations take into account all the $1280$ municipalities. Results are presented for a power law decrease of the weights and a non-linear function $\Psi$ characterized by 3 parameters
(see Materials and Methods for a quantitative comparison of different model variants). Thus, we have here a total of $8$ free parameters: $\omega, A, \zeta_0, d_0, \delta$,  in addition to three for the non-linear function.

\begin{figure}
	\begin{minipage}{\linewidth}
		\centering
		\includegraphics[width=\linewidth]{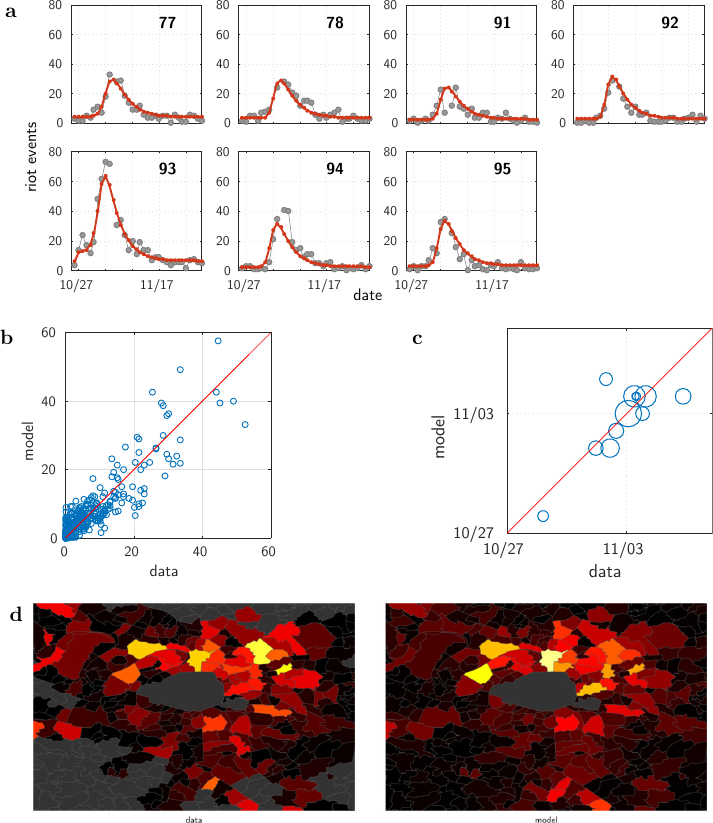}
		\caption{Results: \^Ile-de-France région, model calibration at the scale of municipalities. (a) Time course of the riots: data (dots) and model (continuous curves) -- results presented here are aggregated by d\'epartement. (b) Total number of events, model vs. data. Each dot represents one municipality. In order to compute the total sum of events in the data, missing values were filled using linear interpolation. (c) Temporal unfolding, model vs. data. Date (unit=day) of the maximum rioting activity, shown for the $12$ most active municipalities (those with more than $30$ events). Each circle has a diameter proportional to the size of the reference population of the corresponding municipality. The red lines depict the identity diagonal line. (d) Geographic map of the total rioting activity. Data (left) vs Model (right), shown for the inner suburb of Paris (the ``petite couronne'', d\'epartements 92, 93 and 94). For each municipality, the color codes the total number of events (the warmer the larger, same 
scale for both panels; gray areas: data not available).}
			% The maps have been generated with the Mapping toolbox of the MATLAB\cite{MATLAB} software making use of the Open Street Map data \copyright OpenStreetMap contributors (\url{https://www.openstreetmap.org/copyright}).
		\label{fig:fit_com_idf}
	\end{minipage}
\end{figure}

\begin{figure}
	\begin{minipage}{\linewidth}
		\centering
		\includegraphics[width=\linewidth]{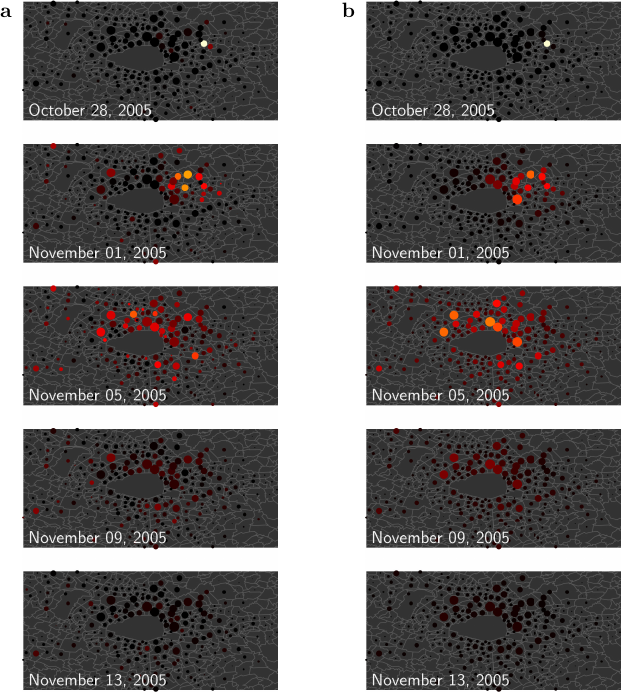}
		\caption{Timeline of the riots in Paris area (\^Ile-de-France). The rioting activity is shown every $4$ days, starting on the day following the triggering event (top). (a) data smoothed by making use of the fitted values given by the single site models (see Supplementary Video 1 for the full dynamics). (b) dynamics generated by the global model (see Supplementary Video 2). The map shows the municipality boundaries, with Paris at the center For each municipality under police authority, a circle is drawn with an area proportional to the size of the corresponding reference population. The color represents the intensity of the rioting activity: the warmer the color, the higher the activity. Figure best viewed magnified in the electronic version.}
			% The maps have been generated with the Mapping toolbox of the MATLAB\cite{MATLAB} software making use of the Open Street Map data \copyright OpenStreetMap contributors (\url{https://www.openstreetmap.org/copyright}).} 
		\label{fig:com_idf_timeline}
	\end{minipage}
\end{figure}

Figures~\ref{fig:fit_com_idf}, \ref{fig:com_idf_timeline} and the Supplementary Video 2 illustrate the main results. Figure~\ref{fig:fit_com_idf} compares the model and the data on four aspects: time course in each d\'epartement (a), amplitude of the events (b), date at which the number of events is maximum (c), and spatial distribution of the riots (d). The global model with a single shock correctly  reproduces the up-and-down pattern at each location, as illustrated on Fig.~\ref{fig:fit_com_idf} at d\'epartement scale. One can note the preservation of the smooth relaxation at each site, despite the influence of other (still active) sites. This can be understood from the SIR dynamics: at a given location, the relaxation term ($-\omega \lambda_k$) dominates when there is no more enough susceptible individuals, so that the local dynamics becomes essentially independent of what is occurring elsewhere. Quite importantly, these local patterns occur at the correct times. One sees that the date of maximum activity 
spreads over several days and varies across locations, which reflects the propagation of the riot.

On the Supplementary Video 2 one can see the wave generated by the model. Figure~\ref{fig:com_idf_timeline}b shows a sketch of this wave as a timeline with one image every $4$ days -- which corresponds to the timescale found by the parameter optimization, $1/\omega \sim 4$ days. 
%% NEW NEW
For comparison, we show side by side, Fig.~\ref{fig:com_idf_timeline}a, the timeline built from the data which have been smoothed making use of the single site fits. %% end NEW NEW
One can see the good agreement, except for few locations where the actual rioting activity occurs earlier than predicted by the global model. A most visible exception is Argenteuil municipality (north-west of Paris on the map, see Fig.~\ref{fig:com_idf_timeline}, second images from the top), where the Minister of Interior made a speech (October 25) perceived as provocative by the banlieues residents. This could potentially explain the faster response to the triggering event. 

The calibrated model gives the time course of the expected number of events in all municipalities, including those under ``gendarmerie'' authority for which we do not have any data. For all of the later, the model predicts a value remaining very small during all the studied period, except for one, the municipality of Fleury-M\'erogis (see Fig.~\ref{fig:fit_com_idf}d, South of the map). Remarkably, searching in the media coverage, we found that a kindergarten has been burnt in that municipality at that period of time (Nov. 6). 

\subsection*{Fitting the data: the wave across the whole country}

\begin{figure}[!h]
	\begin{minipage}{\linewidth}
		\centering	
		\includegraphics[width=\linewidth]{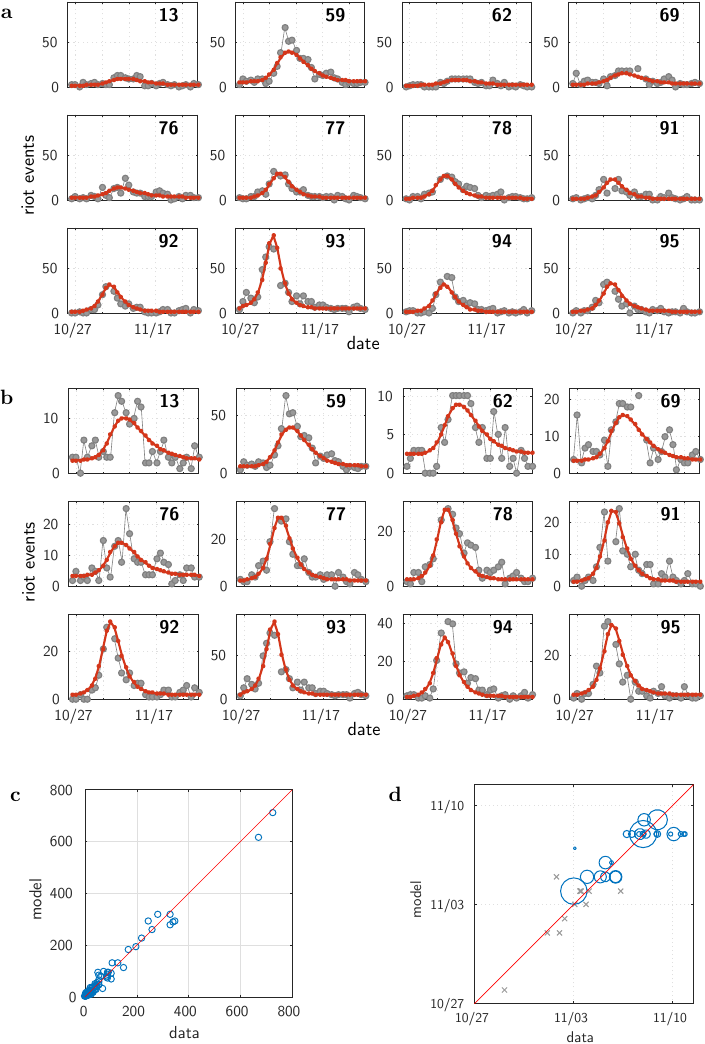}
		\caption{Results: All of France, model calibration at the scale of the d\'epartements.
			(a) Time course of the riots in France: data (dots) and model (continuous curves). Only the $12$ most active d\'epartements are shown. The plots share a common scale for the number of events. (b) Same as (a), but with relative scales. (c) Total number of events. (d) Temporal unfolding (date when the number of riot events reaches its maximum value), shown for the d\'epartements having more than $60$ events. Each blue circle has a diameter proportional to the reference population of the corresponding d\'epartements. The gray crosses remind the plot locations of the major municipalities of \^Ile-de-France shown in Fig.~\ref{fig:fit_com_idf}c. The red lines depict the identity diagonal line.}
		\label{fig:fit_dpt_summary}
	\end{minipage}
\end{figure}

\begin{figure}
	\begin{minipage}{\linewidth}
		\centering
		\includegraphics[width=\linewidth]{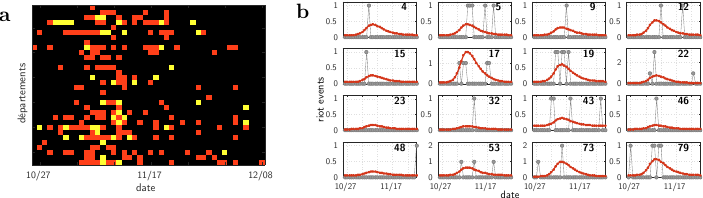}
		\caption{Minor sites: Even where the number of events was very small, one can detect the riot wave. 
			(a) Raster plot of the activities of the 32 départements having at most 2 events each day (black: no event, red: 1 event, yellow: 2 events; the départements are ordered by their ID number, as on Supplementary Fig. S4). 
			(b) Data together with the predictions of the epidemiological model -- the one of Fig.~\ref{fig:fit_dpt_summary}. 
			For ease of presentation, plots only shown for a subset, the 16 with the smallest total number of events on the period (see Supplementary Fig. S4 for the full set of 32 départements).}
		\label{fig:linear_nodip_dpt_fr_minsites}
	\end{minipage}
\end{figure}

We now show that the same model reproduces the full dynamics across the whole country. We apply our global model considering each one of the d\'epartements of metropolitan France (except Corsica and Paris, hence $93$ d\'epartements) as one homogeneous site -- computing at municipality scale would be too demanding (more than $36,000$ municipalities). The Materials and Methods section details the comparison between various model options. We present here the results for the model version making use of the linear approximation, with 
9 free parameters: $\omega, A, \zeta_0, d_0, \xi$, the same susceptibility $\beta$ everywhere except for three different values, for the départements $13$, $62$ and $93$. As for the wave around Paris, the resulting fit is very good, as illustrated on Fig.~\ref{fig:fit_dpt_summary} (see also Supplementary Fig. S3). Figures~\ref{fig:fit_dpt_summary}a and \ref{fig:fit_dpt_summary}b show the results for the $12$ most active d\'epartements. Figure~\ref{fig:fit_dpt_summary}c compares model and data on the total number of events, and Fig.~\ref{fig:fit_dpt_summary}d on the date of the maximum activity. For the latter, the data for the \^Ile-de-France municipalities (Fig.~\ref{fig:fit_com_idf}c) are reported. One sees that the wave indeed spread over all France, with the dynamics in Paris area essentially preceding the one elsewhere.\\
Remarkably, one can see the effect of the riot wave even where few rioting events have been recorded. 
The data exhibit a concentration of (weak) activities (Fig.~\ref{fig:linear_nodip_dpt_fr_minsites}a), a pattern which would not be expected in case of independent random events. The epidemiological model predicts these minor sites to be hit by the wave, with a small amplitude and at the correct period of time. This is apparent on Fig.~\ref{fig:linear_nodip_dpt_fr_minsites}b and can be shown to be statistically significant (see Materials and Methods and Supplementary Fig.~S4 for more details).\\
Finally, we validate here the hypothesis that it is the number, and not the proportion, of individuals (susceptible individuals, rioters) that matters. The very same model, but with densities and not numbers, yields a much less good fit (see Materials and Methods). This comes as a quantitative confirmation of the hypothesized bandwagon effect, in line with previous literature\cite{Granovetter_1978,Raafat_etal_2009}.

\section*{Discussion}
Studying the dynamics of riot propagation, a dramatic instance of large-scale social contagion, is difficult due to the scarcity of data. The present work takes advantage of the access to detailed national police data on the 2005 French riots that offer both the timescale of the day over a period of 3 weeks, and the geographic extension over the country. These data exhibit remarkable features that warrant a modeling approach. We have shown that a simple parsimonious epidemic-like model combining contagion both within and between cities, allows one to reproduce the daily time course of events, revealing the wave of contagion. The simplest model version with only 6 parameters already accounts for the wave very well, and more elaborated versions with about 10 parameters account for even finer details of the dynamics. A crucial model ingredient is the choice of a single sociological variable, taken from the census statistics as a proxy for calibrating the size of the susceptible population. It shows that the 
wave propagates in an excitable medium of deprived neighborhoods. 

It is interesting to put in contrast the results obtained here with a model where homogeneous weights are independent of the geographic distance, which we can consider as a null hypothesis model with regards to the geographic dependency. As discussed in Materials and Methods and illustrated on Supplementary Fig. S5, such a  hypothesis fails to produce a wave, and, more unexpectedly, cannot account for the amplitudes of the riot. This confirms that diffusion by geographic proximity is a key underlying mechanism, and points towards the influence on the riots breadth  of the concentration of urban areas with a high density of deprived neighborhoods (as it is the case for the départements of Île-de-France, 77, 78, 91, 92, 94 and 95). Thus, one can conclude that, having the outbreak location surrounded by a dense continuum of deprived neighborhoods made the large-scale contagion possible.

What lesson on human behavior can we draw from our analysis? 
First, as we just indicated, ``geography matters''\cite{Goldenberg_Levy_2009}: despite the modern communication media, physical proximity is still a major feature in the circulation of ideas or behaviors, here of rioting. Second, strong interpersonal ties are at stake for dragging people into actions that confront social order. The underlying interpretation is that interpersonal networks are relevant for understanding riot participation. Human behavior is a consequence not only of individuals' attributes but also of the strength of the relation they hold with other individuals\cite{Emirbayer_Goodwin_1994}. 
Strong interpersonal connections to others who are already mobilized draw new participants into particular forms of collective action such as protest, and identity (ethnic or religious based) movements\cite{Gould_2003,Walgrave_Wouters_2014}. Third, concentration of socio-economic disadvantage facilitates formation  of a sizeable group and therefore involvement in destruction: the {\em numbers} of rioters in the model (rather than proportions) can be interpreted as an indirect indication of risk assessment before participating in a confrontation with the police\cite{Carter_1987,Gross_2011}. From this viewpoint, rioters seem to adopt a rational behavior and only engage in such event when their number is sufficient.

The question of {\em parsimony} is of the essence in our modeling approach: an outstanding question was to understand whether a limited number of parameters might account for the observed phenomena at various scales and in various locations. We answer this question  positively here, thus revealing the existence of a general mechanism at work: {\em general}, since (i) the model is consistent with what has occurred at each location hit by the riot, and (ii) a similar up-and-down pattern is observed for the US ethnic riots in different cities\cite{Burbeck_etal_1978}, suggesting that this process is indeed common to a large class of spontaneous riots. The wave we have exhibited has a precise meaning supported by the mathematical analysis. Indeed, it is generated by a {\em single} triggering event, with a mechanistic-like dynamics giving to the ensemble of local riots a status of a single global episode occurring at the scale of the country, with a well-defined timescale for the propagation. 

Whether an initially local riot initiates a wave, and, if so, what is its geographic extension, depend on conditions similar to those at work for disease propagation: a high enough density of susceptible individuals, a suitable contact network and large enough susceptibilities. The 2005 riot propagation from place to place after a single shock is reminiscent of the spreading of other riots,  such as the one of food riots  in the late eighteen century in the UK\cite{Bohstedt_Williams_1988}, or of local propagations during the week of riots in the US in reaction to Martin Luther King assassination. The latter series of riots has been coined as a ``wave within a wave'', the larger wave corresponding to the series of US ethnic riots from 1964 to 1971\cite{Myers_2010}. However, this larger `wave' does not appear to be of the same nature as the traveling wave discussed here. Indeed, first, most of the riots in this long time period in the US have each their own triggering event. Second, these events are separated 
one from another by large times gaps and are therefore discontinuous whereas we describe the continuous epidemiological spreading by a wave. To discuss such series of riots as these over long period of times, we note that there is no conceptual difficulty in extending our model to larger timescales -- by adding a weak flux from recovered to susceptible individuals, and by dealing with several shocks --, although one would also have to take into account group identity changes and the effect of policies on structural characteristics of cities. However, the main issue here is rather the access to a detailed set of data.

In any case, the modeling approach introduced here provides a generative framework, different from the statistical/econometric approach, that may be adapted to the detailed description of the propagation of spontaneous collective uprisings from a main triggering event -- notably, the interaction term in our SIR model can be modified to include time delays (time for the information to travel), to take into account time integration of past events, or to be also based on non-geographic criteria (e.g. cultural, ethnic, socio-economic similarity features). We believe that such extensions will lead to interesting developments in the study of spreading of social behaviors.

\section*{Materials and Methods}

\subsection*{French administrative divisions} 
The three main French administrative divisions are: the ``commune'', which we refer to as municipality in the paper (more than $36,000$ communes in France); at a mid-level scale the ``d\'epartement'', somewhat analogous to the English district ($96$ d\'epartements in Metropolitan France, labeled from $1$ to $95$, with $2A$ and $2B$ for Corsica); the ``r\'egion'' aggregating several neighboring d\'epartements ($12$ in Metropolitan France, as of 2016, excluding Corsica). At a given level, geographic and demographic characteristics are heterogeneous. The typical diameter of a d\'epartement is $\sim 100$ km, and the one of a région, $\sim 250$ km.\\
There are two national police forces, the ``police'' and the ``gendarmerie'' (a civilian like police force reporting to the ministry of Interior whose agents have a military status in charge of policing the rural parts of the country). Most urbanized areas (covering all municipalities with a population superior to 20000) are under police authority. The more rural ones are under gendarmerie authority. The available data for the present study only concern the municipalities under police authority, except Paris, for which we lack data (but was not much affected by the riots).\\ 
The full list of the municipalities is available on the French government website, 
\url{https://www.data.gouv.fr/fr/datasets/competence-territoriale-gendarmerie-et-police-nationales/}.

\subsection*{Dataset}
\textbf{\em From the source to the dataset.}
The present analysis is based on the daily crime reports of all incidents of civil unrest reported by the French police at the municipalities under police authority\cite{Cazelles_etal_2007} (see above). 
We have been working with this raw source, that is the set of reports as transmitted by the local police departments, before any formatting or recoding by the national police statistical unit. 
The daily reports are written in natural language, and have been encoded to allow for statistical treatment. From the reports we selected only facts related to urban violence. Some facts are reported more than once (a first time when the fact was discovered, and then one or two days later e.g. if the perpetrators have been identified). We carefully tried to detect and suppress double counting, but some cases may have been missed. 

In the police data, incidents in relation with the riots are mostly cases of vehicles set on fire (about 70\%), but also burning of public transportation vehicles, public buildings,  of waste bins, damages to buses and bus shelters, confrontations between rioters and police, etc. Facts have been encoded with the maximum precision: day and time of the fact, who or what was  the target, the type of damage, the number and kind of damaged objects and the number and quality of persons involved,  whenever these details are mentioned in the report. In the present work, as explained in the main text, from these details, we compute a daily number of events per municipality. We generated a dataset for these events (with a total of $6877$ entries concerning $853$ municipalities). 

There are a few missing or incomplete data -- notably in the nights when rioting was at its maximum, as the police was overwhelmed and reported only aggregated facts, instead of details city per city. 
Note that if one has, e.g., a number of burnt cars only at the département scale, one cannot know what is the corresponding number of events, since each one of these events at municipality scale corresponds to an unknown number of burnt cars. Hence if for a particular day, the police report gives the information only aggregated at département scale, one cannot even make use of it when modeling at this scale. Quantitatively, for the analysis of the events in \^Ile-de-France, working with the  $462$ municipalities under police authority, there is $1.6\%$ of missing data ($334$ out of $462\times 44$days $=20,328$ data values). For the analysis at the scale of the  $93$ départements, there is $0.2\%$ of missing values ($10$ out $93\times 44=4092$). 

\noindent \textbf{\em Datasets for future works.} 
In addition, we have also built two other datasets. From the same police source, we built a dataset for the arrests ($2563$ entries), that in forthcoming work will serve to characterize the rioters as well as to investigate the deterrent effect of arrests on the riot dynamics.  In future work we also plan to explore the rioting events beyond the sole number of events as studied here. One expects to see what has been for instance the role, if any, of curfews and other deterrent effects (in space and time). We also plan to study whether or not the intensity per event smoothly relaxes like the number of events itself. From both local newspapers and national TV and radio broadcasts, we built a specific dataset of media coverage. In ongoing work we extend our modeling framework by considering the coupling between the dynamics of riot events and the media coverage.

\subsection*{Background activity}
The rioting activity appears to be above a constant level which most likely corresponds to criminal activities (an average of an order of $100$ vehicles are burnt every day in France, essentially due to criminal acts not related to collective uprising). Since this background activity has the same level before and after the riot, we assumed that the dynamics of the riot and of the criminal activities are 
independent. In addition, we also considered alternative models where the background activity and the rioting activity would be coupled, the background activity being considered as an equilibrium state, and the riot as a transient excited state. Such models would predict an undershooting of the activity just after the end of the riot -- more exactly a relaxation with damped oscillations --, but the data do not exhibit such behavior 

For the fits, the background activity $\lambda_b$ is taken as the mean activity over the last two weeks in our dataset (November 25 to December 8), period that we can consider as the tail of the data, for which there is no longer any riot activity (see Fig.~\ref{fig:single_site_fits}). For the sites with a non zero number of events in the tail, we observe that this baseline rate is proportional to the size of the reference population chosen for calibrating the size of the susceptible population. For the sites where the number of events in the tail is either zero or unknown (which is the case for a large number of small municipalities, in particular the ones under gendarmerie authority), one needs to give a non zero value to the corresponding baseline rate in order to apply the maximum likelihood method (see below). We estimated it from the size of the reference population (set as 1 when it is 0), using the latter  proportionality coefficient (with a maximum value of $\lambda_b$ set to one over the length of 
the tail, \textit{ie} $1/14$).\\
Statistical tests of the Poisson hypothesis are provided below, paragraph {\it Poisson noise assumption, Stationary tails statistics}.

\subsection*{Epidemiological modeling: Single site model}
We detail here the compartmental SIR model when applied to  
each site {\em separately} (each municipality, or, after aggregating the data, for each d\'epartement). Let us consider a particular site (we omit here the site index $k$ in the equations). 
At each time $t$ there is a number $S(t)$ of individual {\it susceptible} to join the riot, and $I(t)$ of {\it infected} individuals (rioters). Those who leave the riots become  {\it recovered} individuals.
Since we assume that there is no flux from {\it recovered} to {\it susceptible}, we do not have to keep track of the number of recovered individuals. Initially, before a triggering event at some time $t_0$ occurs, there is a certain number $S_0 > 0$ of susceptible individuals but no rioters, that is, $I(t) = 0$ for all $t\leq t_0$. At $t_0$ there is an exogenous shock and the number of rioters becomes positive $I(t_0)=I_0 >0$. From there on, neglecting fluctuations, the numbers of rioters and of susceptible individuals evolve according to the following set of equations:
\begin{subequations}
	\begin{empheq} 
	[left=\empheqlbrace]{align} 
	\frac{d I(t)}{dt} &= - \,\omega \, I(t) + S(t) \, P(s\rightarrow i,t) \label{eq_sir1}\\
	\frac{d S(t)}{dt} &= - \,S(t) \, P(s\rightarrow i,t) 
	\label{eq_sir2}
	\end{empheq}
\end{subequations}
Let us now explain this system of equations. 
In Eq. (\ref{eq_sir1}), $\omega$ is the constant rate at which rioters leave the riot. 
The second term in the right hand side of  Eq.~(\ref{eq_sir1}) gives the flux from susceptible to infected as the product of the number of susceptible individuals, times the probability $P(s\rightarrow i,t)$ for a susceptible individual to become infected. The second equation, Eq.~(\ref{eq_sir2}), simply states that those who join the riot leave the subpopulation of susceptible individuals. 

We now specify the probability to join the riot, $P(s\rightarrow i,t)$  (to become infected when in the susceptible state). 
In line with accounts of other collective uprising phenomena\cite{Gross_2011}, testimonies from participants in the 2005 riots suggest a bandwagon effect: individuals join the riot when seeing a group of rioters in action. Threshold decision models\cite{Schelling_1973,Granovetter_1978} describe this herding behavior assuming that each individual has a threshold. When the herd size is larger than this threshold the individual joins the herd. Granovetter\cite{Granovetter_1978} has specifically applied such a model to riot formation, the threshold being then the number of rioters beyond which the individual decides to join the riot. Here we make the simpler hypothesis that the probability to join the riot does not depend on idiosyncratic factors, and is only an increasing function of the {\em  total} number of rioters at the location (site) under consideration. It is worth emphasizing that this herding behavior is in contrast with the epidemic of an infectious disease, where contagion typically occurs from 
dyadic interactions, in which case the probability is proportional to the {\em fraction} of infected individuals, $I(t)/S_0$. 
Being a probability, $P(s\rightarrow i,t)$ must saturate at some value (at most $1$) for large $I$, and is thus a non-linear function of $I$. Nevertheless, we will first assume that conditions are such that we can approximate $P(s\rightarrow i,t)$ by its linear behavior: $P(s\rightarrow i,t) \sim \kappa I(t)$ (but note that $\kappa$ does not scale with $1/S_0$) and discuss later a different specification for this term.

Given the assumption $\lambda(t) = \alpha I(t)$, it is convenient to define
\begin{equation}
\sigma(t) = \alpha S(t)
\end{equation}
so that the riot dynamics at a single (isolated) site is described by:
\begin{subequations}
	\begin{empheq}[left=\empheqlbrace]{align}
	\frac{d \lambda(t)}{dt} &= - \omega \, \lambda(t) + \beta \, \sigma(t) \, \lambda(t) \\
	\frac{d \sigma(t)}{dt} &= - \beta \, \sigma(t) \, \lambda(t) 
	\end{empheq}
\end{subequations}
where $\beta \equiv \frac{\kappa}{\alpha}$. 
Initially $\lambda = 0$, which is a fixed point of this system of equations. With $\sigma(t_0)=\sigma_0 >0$, the riot starts after the shock  
if the reproduction number\cite{Diekmann_Heesterbeek_2000} $R_0\equiv \beta \sigma_0/\omega=\kappa S_0/\omega$ 
is greater than $1$. In such a case, from $t=t_0$ onward, the number of infected individuals first increases, then goes through a maximum and eventually relaxes back towards zero. 
Because $\kappa$ is {\it not} of order $1/S_0$, this condition seems too easy to satisfy: at any time, any perturbation would initiate a riot. One may assume that the particular parameter values allowing one to fit the data describe the state of the system at that particular period. Previous months and days of escalation of tension may have led to an increase in the susceptibility $\kappa$, or in the number of susceptible individuals $S_0$. 

\subsection*{Epidemiological modeling: Non local contagion} We give here the details on the global SIR model,  with interactions between sites. We have a discrete number $K$ of sites, with homogeneous mixing within each site, and interactions between sites. At each site $k$, there is a  number $S_k$  of ``susceptible'' individuals, $I_k$ of ``infected'' (rioters), and $R_k$  of ``recovered'' individuals. As  above, there is no flux from recovered to susceptible (hence we can ignore the variables $R_k$), and individuals at site $k$ leave the riot at a constant rate $\omega_k$. Assuming homogeneous mixing in each site, the dynamics is given by the following set of equations:
\begin{subequations}
	\label{meth_sir_nonloc_pop} 
	\begin{empheq}[left=\empheqlbrace]{align}
	\frac{d I_k(t)}{dt} &= - \,\omega_k \, I_k(t) + S_k(t) \, P_k(s\rightarrow i,t)\\
	\frac{d S_k(t)}{dt} &= - \,S_k(t) \, P_k(s\rightarrow i,t) 
	\end{empheq}
\end{subequations}
with the initial conditions $t<t_0 \; I_k(t) = 0,\; S_k(t)=S_{k 0} > 0$, and at $t=t_0$, a shock occurs at a single location $k_0$, $I_k(t_0)=I_0 >0$. 
In the above equations, $\omega_k$ is the local recovering rate, and $ P_k(s\rightarrow i,t)$ is the probability for a $s$-individual at location $k$ to become a rioter at time $t$. 

We now write the resulting equations for the $\lambda_k$. We assume the rioting activity to be proportional to the number of rioters:
\begin{equation} 
\lambda_k(t)= \alpha I_k(t)
\label{meth_lai}
\end{equation} 
Note that different hypothesis on the dependency of $\lambda_k$ on $I_k$ could be considered. For instance we tested  $\lambda_k \sim( I_k)^q$ with some exponent $q$ coming as an additional free parameter. In that case, the  optimization actually gives that $q$ is close to $1$. 

Multiplying each side of (\ref{meth_sir_nonloc_pop}) by $\alpha$, one gets
\begin{subequations}
	\label{meth_sir_nonloc}
	\begin{empheq}[left=\empheqlbrace]{align}
	\frac{d \lambda_k(t)}{dt} &= - \,\omega_k \,  \lambda_k(t) + \sigma_k(t) \, P_k(s\rightarrow i,t)\\
	\frac{d \sigma_k(t)}{dt} &= - \,\sigma_k(t) \, P_k(s\rightarrow i,t)
	\end{empheq}	
\end{subequations}
where as before we introduce $\sigma_k(t)=\alpha S_k(t)$. 
Taking into account the hypothesis on the linear dependency of the number of event in the number of rioters, (\ref{meth_lai}), we write $P(s\rightarrow i,t)$ directly in term of the $\lambda$s:
\begin{equation} 
P_k(s\rightarrow i,t) = \Psi_k(\Lambda_k(t)) 
\label{meth_psipsi}
\end{equation} 
where  $\Lambda_k(t)$ is the activity  ``seen'' from site $k$ (see main text):
\begin{equation} 
\Lambda_k(t) \equiv \sum_j W_{kj} \lambda_j(t) 
\label{meth_Lambda}
\end{equation} 
where the weights $W_{kj}$ are given by a decreasing function of the distance $\text{dist}(k,j)$ between sites $k$ and $j$: $W_{kj}= W(\text{dist}(k,j))$ (see below). The single site case is recovered for  $W_{kj}=\delta_{k,j}$.

In the linear approximation,
\begin{equation}
\Psi_k(\Lambda)  = \beta_k \, \Lambda,
\end{equation}
in which case one gets the set of equations
\begin{subequations}
	\label{si_sir_nonloc_lin}
	\begin{empheq}[left=\empheqlbrace]{align}
	\frac{d \lambda_k(t)}{dt} &= - \,\omega_k \,  \lambda_k(t) + \,\beta_k\,\sigma_k(t) \, \sum_j W_{kj} \lambda_j(t) \\
	\frac{d \sigma_k(t)}{dt} &= - \,\beta_k\, \sigma_k(t) \, \sum_j W_{kj} \lambda_j(t).  
	\end{empheq}
\end{subequations}
The form of these equations is analogous to the ones of the original 
distributed contacts continuous spatial SIR model~\cite{Kendall_1965} (see below) but here with a discrete set of spatial locations.\\
In the whole paper, we take a site independent value of the recovering rate, $\omega_k=\omega$ for every site $k$. Similarly, the susceptibility is chosen site-independent, $\beta_k=\beta$, except for some variants where a few sites are singularized, see section \emph{Results, details: All of France, département scale}, below.\\
In the non-linear case, we choose parameters for $\Psi_k(\Lambda)$ in order to have a function (i) being zero when there is no rioting activity; (ii) which saturates at a value (smaller or equal to $1$) at large argument; (iii) with a monotonous increasing behavior giving a more or less pronounced threshold effect (e.g. a sigmoidal shape). This has to be done looking for the best compromise between quality of fit and number of parameters (as small as possible). We tested several sigmoidal functions. 
For the fit of the Paris area at the scale of the municipalities, we made use of a variant with a strict threshold:
{\small
	\begin{subequations}
		\label{meth_psi_thr}
		\begin{empheq}[left=\empheqlbrace]{align}
		\Lambda \leq \Lambda_{c k},\,\,& \Psi_k(\Lambda) = 0\\
		\Lambda > \Lambda_{c k},\,\,& \Psi_k(\Lambda) = \eta_k\; \left(1- \exp{-\gamma_k\;(\Lambda-\Lambda_{c k})}\right)
		\end{empheq}
	\end{subequations}
}
The fit being done with site-independent free parameters, this function thus contributes to three free parameters, $\Lambda_c, \eta$ and $\gamma$. 

\subsection*{Choice of the weights}
The best results are obtained for two options. One is a power law decay with the distance:
\begin{equation} 
W_{kj} = \left(1+\text{dist}(k,j)/d_0\right)^{-\delta}
\label{eq_wkj}
\end{equation} 
where $ \text{dist}(k,j)$ is the distance between site $k$ and site $j$ (see below for its computation). 
The second option is the sum of an exponential decay and of a constant term 
\begin{equation} 
W_{kj} = \xi\;+\; (1-\xi) \exp\left(-\text{dist}(k,j)/d_0\right) 
\label{eq_wkjbis}
\end{equation}
In both cases we normalize the weights so that for every site $k$, $W_{k k}=1$. Taking site-independent free parameters, both cases give two free parameters, $d_0$ and $\delta$ for the choice (\ref{eq_wkj}), $d_0$ and $\xi$ for the choice (\ref{eq_wkjbis}).

\subsection*{Distance-independent null hypothesis model}
In order to test for the possible absence of geographic dependency in the contagion process, we take as a ``null hypothesis'' model a version of our model where the weights $W_{kj}$ do not depend on the distance between sites. In this version, a given site is concerned by what is happening at its own location, and in an equally fashion by what is happening elsewhere. Mathematically, we thus consider the following weights:
\begin{equation}
W_{kj} = 
\begin{cases}
1, & \text{if}\ k=j \\
\xi, & \text{otherwise}
\end{cases}
\end{equation}
where $\xi$ is a constant term to be optimized as a free parameter. Apart from the choice of the weights, the model is the same as the one corresponding to the results shown on  Fig.~\ref{fig:fit_dpt_summary}. Optimization is done over all the 8 free parameters. The results for this model are shown on Supplementary Fig.~S5, to be compared with Fig.~\ref{fig:fit_dpt_summary}. 
As one would expect, this model does not generate any wave: following the shock, all the riot curves happen to peak at the same time. Remarkably, assuming no geographic effect in the interaction term does not simply affect the timing of the events: the model also fails to account for the amplitudes of the rioting activities (see Supplementary Fig. S5b). \\ 
For what concerns the statistical significance, as expected from the comparison between the figures, the distance-independent null hypothesis model is far worse than the model with geographic dependency, despite the fact that it has one less free parameter. Making use of the Akaike Information Criterion\cite{Akaike_1974} (AIC), the difference in AIC is $\Delta\text{AIC} = -438$, which corresponds to a relative likelihood\cite{Burnham_Anderson_2003} of $9.0e-96$. A similar conclusion is obtained from the BIC criterion\cite{Schwarz_1978}.\\
These results thus clearly underline the need for the interaction term to depend on the distance, supporting the view of a local contagion process.

\subsection*{Classic SIR model with densities}
A more direct application of the classic SIR model as used for infectious diseases would have lead to consider equations for {\it densities} of agents (instead of numbers of agents). In the linear case, this leads to the following equations for the $\lambda$s and $\sigma$s:
\begin{subequations}
	\label{si_sir_norm}
	\begin{empheq}[left=\empheqlbrace]{align}
	\frac{d \lambda_k(t)}{dt} &= - \,\omega \,  \lambda_k(t) + \,\beta\,\sigma_k(t) \, \sum_j W_{kj} \frac{\lambda_j(t)}{N_j} \\
	\frac{d \sigma_k(t)}{dt} &= - \,\beta\, \sigma_k(t) \, \sum_j W_{kj} \frac{\lambda_j(t)}{N_j}.  
	\end{empheq}
\end{subequations}
where here $\beta=\kappa/\zeta_0$, and $N_j$ is the size of the reference population at location $j$. These equations should be compared with Eq.~(\ref{si_sir_nonloc_lin}). Note that the weights being given by (\ref{eq_wkjbis}), the dependency in the population size $N_j$ cannot be absorbed in the weights. 

Fitting this model with densities to the data leads to a much lower likelihood value compared to the model presented here. In the case of the fit of the whole dynamics at the scale of the départements, the difference in AIC is $\Delta\text{AIC} = -101$, which corresponds to a relative likelihood of $9.6e-23$.

\subsection*{Links to the original spatially continuous SIR model}
In the case of the linear approximation, the meta-population SIR model that we have introduced leads to the set of equations (\ref{si_sir_nonloc_lin}) of a type similar to the space-continuous non local (distributed contact) SIR model. With a view to describe the spreading of infections in spatially distributed populations, Kendall\cite{Kendall_1957} introduced in 1957 this non-local version of the Kermack-McKendrick SIR model in the form of space-dependent integro-differential equations. Omitting the recovered population $R$, the system in the $S,I$ variables reads:

\begin{subequations}
	\footnotesize
	\label{si_kendal}
	\begin{empheq} 
	[left=\empheqlbrace]{align}
	\frac{dI(x,t)}{dt} &= -\omega I(x,t) + \beta S(x,t) \int K(x,y) I(y,t) dy \\
	\frac{dS(x,t)}{dt} &= -\beta S(x,t) \int K(x,y) I(y,t) dy
	\end{empheq}
\end{subequations}
where $x\in\mathbb{R}^N$, with $N=1, 2$, and here $I(x,t)$ and $S(x,t)$ are \emph{densities} of immune and susceptible individuals. In the particular case of dimension $N=1$, and the space is \emph{homogeneous}, meaning here that $K(x,y)$ is of the form $K(x,y)=w(x-y)$, we know\cite{Kendall_1965} that there exist traveling waves of any speed larger than or equal to some critical speed. Furthermore, this critical traveling wave speed also yields the asymptotic speed of spreading of the epidemic\cite{Aronson_1977}. There have been many mathematical works on this system and on various extensions\cite{Ruan_etal_2007,Wang_Wood_2011}. Thus, at least in dimension $N=1$ and for homogeneous space,  this non-local system can generate traveling fronts for the density of susceptible individuals, hence the propagation of a ``spike'' of infected individuals. Although no proof exists in dimension $N=2$, numerical simulations show that the model can indeed generate waves\cite{Bailey_1967,Rodriguez-Meza_2012}, as illustrated 
 by the Supplementary Videos 3 and 4, similar to the way the riot spread around Paris giving rise to the informal notion of a \emph{riot wave}.

However, the model we introduce here is more general and differs from the Kendall model in certain aspects. Indeed, rather than continuous and homogeneous, the spatial structure is discrete with heterogeneous sites. Moreover, the set of equations here (\ref{si_kendal}) corresponds to the linear approximation (\ref{si_sir_nonloc_lin}), whereas our general model involves a non-linear term. The understanding of \emph{generalized} traveling waves and the speed of propagation in this general context are interesting open mathematical problems. 

More work is needed to assess the mathematical properties of the specific family of non local contagion models introduced here, that is defined on a discrete network, with highly heterogeneous populations, and a non-linear probability of becoming infected. 

\subsection*{Date of the maximum}
Figures~\ref{fig:fit_com_idf}c and~\ref{fig:fit_dpt_summary}b show how well the model accounts for the temporal unfolding of the riot activity, thanks to a comparison between model and data of the date when the riot activity peaks at each location. Given the noisy nature of the data, the empirical date of that maximum itself is not well defined. For each site, we estimated this date as the weighted average of the dates of the $3$ greatest values, weighted by those values. We filled in missing data values by linear interpolation.\\
For the contagion around Paris, considering the 12 most active municipalities shown in Fig.~\ref{fig:fit_com_idf}c, the correlation coefficient is $r = 0.80$, $p=0.0017$. 
At the scale of the whole country, Fig.~\ref{fig:fit_dpt_summary}d, considering the départements having more than 60 events, the correlation coefficient is equal $r=0.77$, with a p-value of $p=5.2e-6$. Given the large differences in population size, the weighted correlation is more appropriate for comparing the timing of the riot activities. Using weights equal to the population sizes, this yields a weighted correlation coefficient of $r=0.87$, with a p-value $p<1e-5$ estimated with a bootstrap procedure.

\subsection*{Non-free parameters: Choice of the reference population}
\textbf{\em Populations statistics.}  
For the choice of the reference population, we compared the use of various specific populations, considering cross-linked database that involve age, sex and diploma. The source of these populations statistics is the INSEE, the French national institute carrying the national census (\url{http://www.insee.fr/}). 
For the period under consideration, we used relevant data from 2006 since data from 2005 were not available.\\ 
When applying the model at the scale of d\'epartements, for each d\'epartement the size of a given specific population is computed as the sum of the sizes of the corresponding populations of all its municipalities that are under police authority.\\
\textbf{\em Choice of the reference population.}  
Working at the scale of départements, we found the best log-likelihood when using as reference population the one of males aged between 16 and 24 with no diploma, while not attending school (see Supplementary Fig. S7).
We thus calibrated the susceptible population in (all variants of) the model by assuming that, for each site (municipality or département), its size is proportional to the one of the corresponding reference population.\\
\textbf{\em Influence on the results.} The choice of the reference population has a major influence on the results. We find that an improper choice cannot be compensated by the optimization of the free parameters. As an example when working at the scale of municipalities, compare Supplementary Fig. S2, for which the reference population is the total population, with Fig.~\ref{fig:fit_com_idf}a and \ref{fig:fit_com_idf}b.

\subsection*{Non-free parameters: geographic data}
The geographic data are taken from the collaborative project Open Street Map (\url{http://osm13.openstreetmap.fr/~cquest/openfla/export/}). \\
The distance $\text{dist}(k,j)$ is taken as the one (in km) between the centroid of each site. In the case of the municipalities, the centroid is taken as the geographic centroid computed with QGIS\cite{QGIS_software}. In the case of d\'epartements, the centroid is computed as the weighted centroid (weighted by the size of the reference population) of all its municipalities that are under police authority.\\
Making use of these geographic data, all the maps (Fig. \ref{fig:fit_com_idf}d and \ref{fig:com_idf_timeline}, and Supplementary Videos 1 and 2), have been generated with the Mapping toolbox of the MATLAB\cite{MATLAB} software.

\subsection*{Free parameters: numerical optimization}
The data fit makes use of the maximum likelihood approach\cite{Myung_2003}. Let us call $X = \{x_{k,i}, k=1\ldots K,i=1\ldots 44 \}$ the data, where each $x_{k,i} \in \mathbb{N}$ corresponds to the number of events for the site $k$ at day $i$ ($i=1$ corresponding to October 26, 2005), and let $\theta$ denote the set of free parameters (e.g. $\theta = \{\omega, A, \zeta_0, d_0, \delta, \beta\}$ in the multi-sites linear case). Assuming conditional independence, we have:
\begin{equation}
p(X|\theta) = \prod_k \prod_i p(x_{k,i}|\theta)
\end{equation}
Under the Poisson noise hypothesis, the $x_{k,i}$ are Poisson probabilistic realizations with mean $(\lambda_{k,i}(\theta) + \lambda_{b k})$:
\begin{equation}
p(x_{k,i}|\theta) = \frac{(\lambda_{k,i}(\theta) + \lambda_{b k})^{x_{k,i}}}{x_{k,i}!} \exp\big(-(\lambda_{k,i}(\theta) + \lambda_{b k})\big)
\end{equation}
The log-likelihood, computed over all the sites under consideration and over the whole period (44 days long) for which we have data, thus writes:
\begin{multline}
\ell(\theta|X) = \log p(X|\theta) \\ = \sum_{k,i} \big(-\lambda_{k,i}(\theta) - \lambda_{b k} + x_{k,i} \log (\lambda_{k,i}(\theta) + \lambda_{b k})\big)\\ - \sum_{k,i} \log x_{k,i}!
\end{multline}
Note that the last term in the right hand side does not depend on the free parameters and we can thus ignore it.\\ 
We performed the numerical maximization of the log-likelihood using the interior point algorithm\cite{Byrd_etal_1999} implemented in the MATLAB\cite{MATLAB} \texttt{fmincon} function.

The method developed here allows one to explore the possibility of predicting the future time course of events based on the observation of the events up to some date. Preliminary results indicate that, once the activity has
reached its peak in the Paris area, the prediction in time and space of the riot dynamics for the rest of France becomes quite accurate. 

\subsection*{Results, details: Paris area, municipality scale}
For the results illustrated by the figures in the paper, we give here the free parameters numerical values obtained from the maximum likelihood method in the case of the fit at the scale of municipalities in \^Ile-de-France. Note that this optimization is computationally demanding: it requires to generate a large number of times (of order of tens of thousands) the full dynamics ($44$ days) with $2560$ ($2\times 1280$) coupled equations. For the choice of the function $\Psi$, we tested the linear case and several non-linear choices. Results are presented for the non-linear case, the function $\Psi$ being given by (\ref{meth_psi_thr}). We find: 
$\omega = 0.26, A=5.5$;
for the power law decrease of the weights, $d_0 =8. \;10^{-3} km, 
\; \delta= 0.67$; 
$\zeta_0=7.7/N_{max}$, where $N_{max} = 1174$ is the maximum size of the reference populations, the max being taken over all \^Ile-de-France municipalities;  
for the parameters of the non-linear function:
$\eta =0.63, \gamma = 1.27, \Lambda_c=0.06$.\\
In Supplementary Table S1a we provide a summary of the variants that we have explored, together with a comparison according to the AIC criterion.

\subsection*{Results, details: All of France, d\'epartement scale}
We detail here the model options and the numerical results 
for the global model, considering each one of the d\'epartements of metropolitan France (except Corsica and Paris, hence $93$ d\'epartements) as one homogeneous site. 
We have thus $186$ $(2 \times 93)$ coupled equations with 6 to 12 free parameters, depending on the choice of the function $\Psi$ and of the number of specific susceptibilities, see below and Supplementary Table S1b.\\
\textbf{\em Outliers.} Looking at the results for different versions, we observe some systematic discrepancy between data and model for three d\'epartements: 93, where the predicted activity is slightly too low and starts slightly too late, and 13 and 62 where it is too high (see Supplementary Fig.~S6b).
Actually, if one looks at the empirical maximum number of events as a function of the size of the reference population used for calibrating the susceptible population, these three d\'epartements show up as outliers: the riot intensity is significantly different from what one would expect from the size of the poor population. 
Outliers are here defined as falling outside the mean $\pm$ 3 standard deviations range when looking at the residuals of the linear regression. If one considers 4 standard deviations from the mean, one only finds the département 13. \\
The cases of 93 and 13 are not surprising. D\'epartement 93 is the one where the riots started, and has the highest concentration of deprived neighborhoods. Inhabitants are aware of this particularity and refer to their common fate by putting forward their belonging to the ``neuf-trois'' (nine-three, instead of ninety three). Events in d\'epartement 13 are mainly those that occurred in the city of Marseille. Despite a high level of criminality, and large poor neighborhoods, the inhabitants consider that being ``Marseillais'' comes before being French, so that people might have felt less concerned. The case of 62 (notably when compared to 59) remains a puzzle for us.\\ 
We have tested the model calibration with variants having possibly one more free parameter for each one of these sites, $\beta_{93}$, $\beta_{62}$ and $\beta_{13}$, allowing for a different value of the susceptibility than the one taken for the rest of France. The quality of fit for different options (a single $\beta$ value, a specific value $\beta_{13}$, and 3 specific values $\beta_{93}$, $\beta_{62}$ and $\beta_{13}$) are shown on Supplementary Table S1b. The best result (with a linear function $\Psi$) is obtained in the case of 3 specific values (with a total of 9 free parameters).\\
\textbf{\em Function $\Psi$.}
We also compared the choice of the linear function with the one of a non-linear function (still with 3 specific $\beta$ values). The best AIC  is obtained with a non-linear $\Psi$, with a total of 12 free parameters. The main qualitative gains with this variant can be seen comparing  Supplementary Fig.~S6c with 
%% NEW NEW
Supplementary Fig.~S6a (which, for ease of comparison, reproduces Fig.~\ref{fig:fit_dpt_summary}b): %% end NEW NEW
a slightly sharper increase at the beginning for every site, and a better value of the maximum activity in département 59. However, the maximum values for départements 94 and 95 are clearly better predicted from the variant with only 9 free parameters. Apart from these main qualitative differences, the fits are essentially equivalent. We thus choose to present the results for the simpler variant with 9 parameters (with a linear function $\Psi$ and three specific $\beta$ values).\\
\textbf{\em Parameterization and results.} Making use of the linear choice for $\Psi$, introducing 
one free parameter for each one of the sites $13, 62$ and $93$, 
and using the weights given by an exponential decrease plus a constant global value, Eq. (\ref{eq_wkjbis}), optimization is thus done over the choice of 9 free parameters:  $\omega, A, \zeta_0, d_0, \xi, \beta, \beta_{13}, \beta_{62}$ and $\beta_{93}$.\\
The numerical values of the free parameters obtained after optimization are as follows: $\omega = 0.41, A=2.6$;  $\zeta_0=190/N_{max}$, where here $N_{max} = 15632$ is the maximum size of the reference populations of all metropolitan d\'epartements; the susceptibility is found to be $\beta=2.\, 10^{-3}$, except for three d\'epartements as explained above. For these three d\'epartements with a specific susceptibility, 
one finds about twice the common value for the d\'epartement $93$ (where riots started), $\beta_{93}/\beta \sim 1.95$, and about half for the d\'epartement $62$ and $13$,  $\beta_{62}/\beta \sim 0.47$ and $\beta_{13}/\beta \sim 0.42$.\\
For the weights chosen with an exponential decrease plus a constant global value, Eq. (\ref{eq_wkjbis}), $d_0= 36$ km, $\xi=0.06$. When using instead the power law decrease, Eq. (\ref{eq_wkj}), one finds that the fit is almost as good. The exponent value is found to be $\delta = 0.80$, which is similar to the value $\delta = 0.67$ found for the fit at the scale of municipalities (restricted to \^Ile-de-France région). Yet, these exponent values are much smaller than the ones, between $1.$ and $2.$, found in the literature  on social interactions as a function of geographical distance\cite{Liben-Nowell_etal_2005,Lambiotte_etal_2008,Goldenberg_Levy_2009}. A small value of $\delta$ means a very slow decrease, a hint to the need of keeping a non zero value at very large distance. This can be seen as another indication that the alternative choice with a long range part, Eq. (\ref{eq_wkjbis}), is more relevant, meaning that both geographic proximity and long range interactions matter.

\subsection*{Minor sites: Comparison with a constant rate null-hypothesis}
The model predicts that the minor sites, that is, those where the number of events is very small (with a level to be chosen, as discussed below) are hit by the wave, with a very small amplitude and at the correct period of time. This can be seen clearly on Fig.~\ref{fig:linear_nodip_dpt_fr_minsites}b and Supplementary Fig. S4a. 
When looking at these figures, one should keep in mind that the model predictions represent the mean values of stochastic Poisson point processes. Thus, for instance, for a given day, and a given site, a value smaller than 1 for the Poisson parameter $\lambda$ means that the most probable situation is no event at all, and we expect, say, $0$ or $1$ event. Yet, one should ask whether the apparent agreement is purely the result of chance. Of course the fit for any one of these minor sites, {\it taken alone}, is not significant. What matters here is the consistency of the global model with the set of activities of all the minor sites. \\
In order to quantitatively evaluate the relevance of the fit even for these minor sites, we confront the model predictions against the predictions of a null model specific for this set of sites. We consider a constant rate null-hypothesis model defined as a Poisson noise model, with a parameter for each site that is constant in time ($\lambda_k(t)=\lambda_k$). This parameter is chosen to be the empirical average number of events over the available period. Supplementary Figure S4b provides the comparison in terms of difference in AIC criterion between the two models. In this comparison, the AIC of the epidemiological model is  obtained  from its calibration over the full set of départements (see Fig.~\ref{fig:fit_dpt_summary}). In the resulting log-likelihood we only keep the terms  that specifically depend on the considered minor sites.\\
We find that, when considering the minor sites as those with at most one event on any single day, the null model is preferred to the epidemiological model. This is not surprising given the  level of noise in our dataset -- compare in particular the presence of a criminal background  not associated with the riots, as discussed above in section \emph{Background activity}. However, when considering the minor sites as those  where the number of events on any day does not exceed a value as low as two, we find that the epidemiological model yields a better account of the activity than the null model (Supplementary Fig. S4b). This is all the more remarkable as these sites have only a small influence on the calibration of the full model. Indeed, their contribution to the global model likelihood is small  compared to the one of the major sites which essentially drive the data fit. As Supplementary Fig. S4b also shows, the gain in AIC increases rapidly when the number of events allowed for defining the minor sites 
increases.

\subsection*{Poisson noise assumption}
\begin{figure}
	\begin{minipage}{\linewidth}
		\centering
		\includegraphics[width=\linewidth]{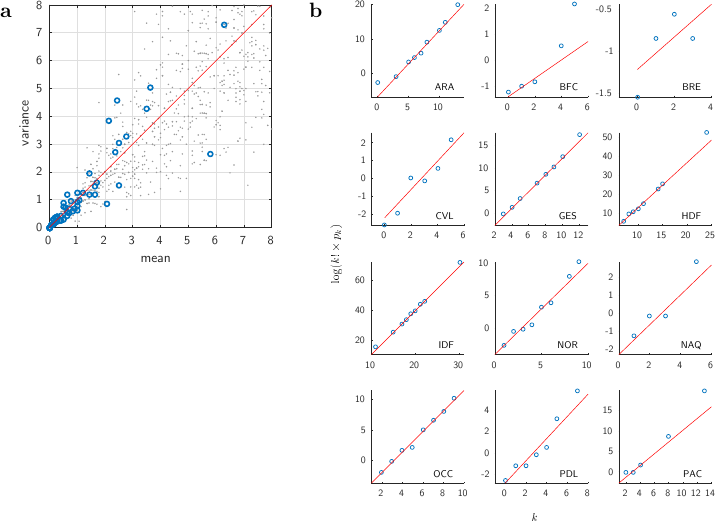}
		\caption{Test of Poisson statistics assumption.
			(a) Mean and variance computed on the tail of the data (taken as the last two available weeks considered as background noise, without rioting activity). Each circle corresponds to one d\'epartement. For comparison, we generated fake Poisson data mimicking the typical values observed in the dataset: the gray dots give the sample mean and variance for $14$ realizations of $1000$ Poisson processes, with true means  randomly generated between $0$ and $10$. 
			(b) For each région, Poissonness plot computed for the tail of the data. See Supplementary Table S2 for more details on the régions.}
		\label{fig:poisson}
	\end{minipage}
\end{figure}

In order to calibrate the model to the data, we assume Poisson statistics, although we do not claim that the underlying process is exactly of Poisson nature. However, it is a convenient working hypothesis for numerical reasons (see above \textit{Free parameters: numerical optimization}). From a theoretical point of view, this choice is a priori appropriate as we deal with discrete values (often very small). In addition, we have seen that the data suggests that a same kind of model is relevant at different scales: this points towards infinitely divisible distributions, such as the Poisson distribution -- the sum of several independent Poisson processes still being a Poisson process.\\
\noindent
\textbf{\em Stationary tails statistics.}
We show here that the statistics of the background activity (data in the tails exhibiting a stationary behavior) are compatible with the Poisson hypothesis. 
Under such a hypothesis, the variance is equal to the mean. Figure~\ref{fig:poisson}a shows that, for each d\'epartement, if we look at the last two weeks, the variance/mean relationship is indeed in good agreement with a Poisson hypothesis.\\ 
As additional support to the Poisson noise property, Fig.~\ref{fig:poisson}b shows a Poissonness plot\cite{Hoaglin_1980_poissonness} for each of the $12$ régions. For completeness, we recall the meaning of a Poissonness plot. One has a total number of observations $n$. Each particular value $x$ is observed a certain number of times $n_x$, hence an empirical frequency of occurrence $n_x / n$. If the underlying process is Poisson with mean $\lambda$, then one must have $\log(x! n_x / n) = - \lambda + x \log(\lambda)$. Thus, in that case, the plot of the quantity $\log(x! n_x / n)$ (the blue circles in  Fig.~\ref{fig:poisson}b) as a function of $x$ should fall along a straight line with slope $\log(\lambda)$ and intercept $-\lambda$ (the red lines on Fig.~\ref{fig:poisson}b).
		
\begin{figure}[!htb]
	\begin{minipage}{\linewidth}
		\centering
		\includegraphics[width=\linewidth]{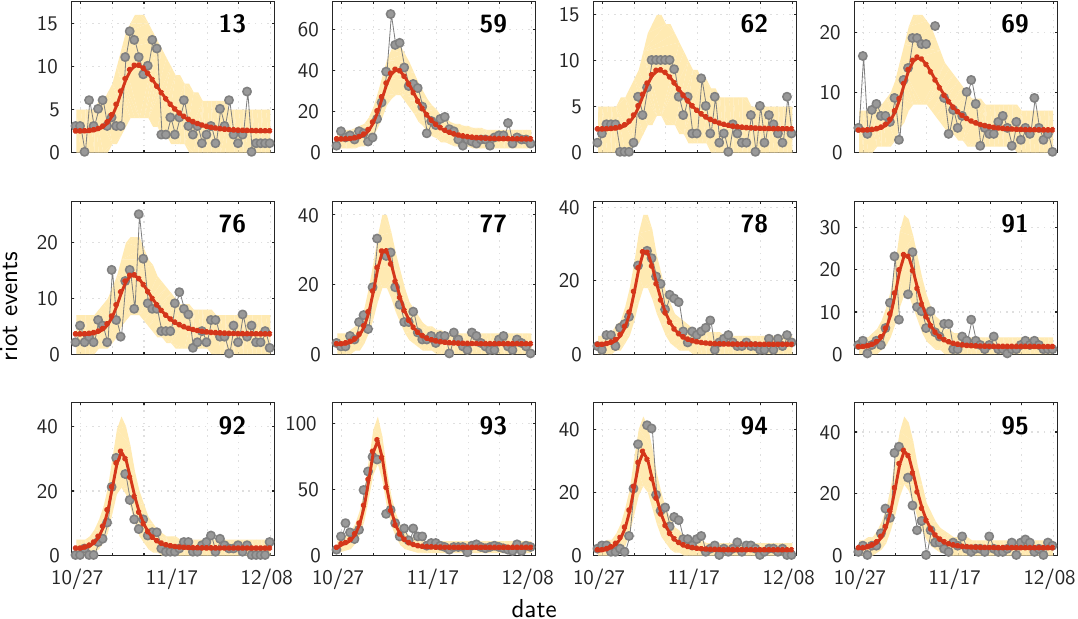}
		\caption{Same figure as Fig.~\ref{fig:fit_dpt_summary}b with highest density regions (HDR). The light orange areas correspond to the 95\% highest density regions.}
		\label{fig:fit_dpt_hdr}
	\end{minipage}
\end{figure}

\begin{figure}[!htb]
	\begin{minipage}{\linewidth}
		\centering
		\includegraphics[width=\linewidth]{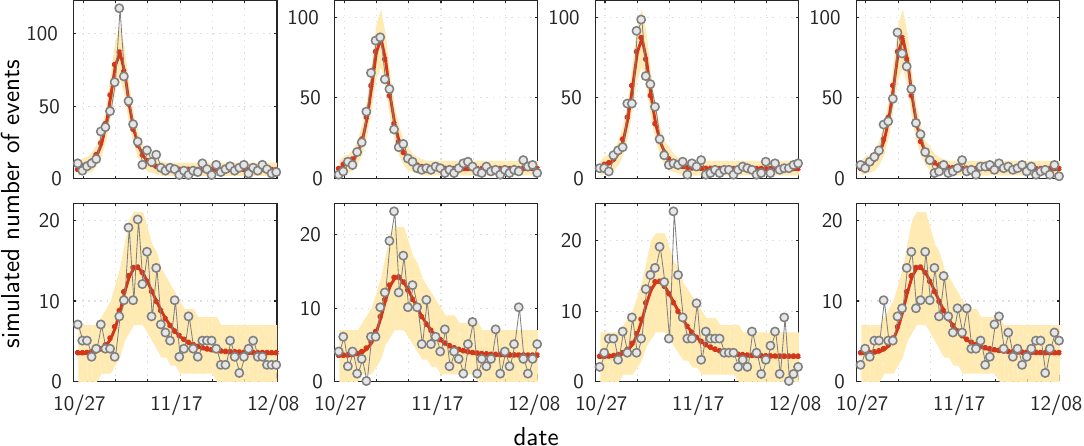}
		\caption{Surrogate data: examples of Poisson samples (open circles) for two examples of rate curves (red curves). These curves are taken from the model fit, top: for département 93, bottom: for département 76. In each case, four different probabilistic realizations of Poisson noise are shown. The light orange areas are the 95\% highest density regions.}
		\label{fig:poisson_real}
	\end{minipage}
\end{figure}

\noindent
\textbf{\em Poisson realizations and Highest Density Regions.}
We remind that we consider the observed data as probabilistic realizations of an underlying Poisson process, whose mean $\lambda(t)$ is the outcome of the model fit. To have a better grasp on the meaning of such data fit, as a complement to  Fig.~\ref{fig:fit_dpt_summary}b, we provide Fig.~\ref{fig:fit_dpt_hdr}.  On this figure we have plotted the 95\% Highest Density Regions~\cite{Hyndman_1996} (HDR, light orange areas) along with the means $\lambda(t)$ (red curves) of the Poisson processes. The rational is as follows. From fitting the model, for each site and for each date, we have a value of $\lambda$. 
If one draws a large number of realizations of a Poisson process with this mean value $\lambda$, one will find that 95\% of the points lie within the corresponding 95\% HDR. More precisely, a 95\% highest density region corresponds to the interval of shortest length with a probability coverage of 95\%~\cite{Hyndman_1996}. \\
For each value of the set of $\lambda$s, outcome of the fit with the global, non local, model, we estimated the corresponding 95\% HDR thanks to a Monte Carlo procedure. These regions are shown as light orange areas on Fig.~\ref{fig:fit_dpt_hdr}. These HDR allow to visualize the expected size of fluctuations (with respect to the mean). 
Next, we look where the actual data points (gray points on Fig.~\ref{fig:fit_dpt_hdr}) lie with respect to the HDR. 
First, one sees that the empirical fluctuations are in agreement with the sizes of the HDRs (qualitatively, the points are spread in the HDRs). Second, remarkably, one finds that the percentage of data points outside the HDR is 9\%, a value indeed close to the expected value $100-95=5\%$ (expected if both the fit is good and the noise is Poisson). 
This however slightly larger value could be due to statistical fluctuations. Yet, a closer look at the plots suggest a few large deviations, such as day 2 in département 69, that might correspond to true idiosyncrasies, cases which cannot be reproduced by the model and show up as 
particular deviations to the ``first order scenario'' we present.\\
In addition to this analysis, to get more intuition on what Poisson fluctuations may produce, we generated artificial data that are Poisson probabilistic realizations given a certain underlying mean $\lambda$. 
Fig.~\ref{fig:poisson_real} present two illustrative cases, where the $\lambda$s are taken as the outcome of the fit for départements 93 and 76. In each case, four different probabilistic realizations are shown.

\newpage

\subsection*{Data availability}
The dataset used for this work is available from the corresponding author on reasonable request, and under the condition of proper referencing.  

\section*{Acknowledgements}
{This paper benefited from the critical reading by several colleagues from various fields, as well as from useful remarks of anonymous referees. All the maps (Fig. \ref{fig:fit_com_idf}d and \ref{fig:com_idf_timeline}, and Supplementary Videos 1 and 2), have been generated thanks to the Open Street Map data \copyright OpenStreetMap contributors (\url{https://www.openstreetmap.org/copyright}). This work received support from: the European Research Council advanced grant ERC ReaDi (European Union's 7th Framework Programme FB/2007-13/, ERC Grant Agreement no 321186 held by H. Berestycki); the CNRS interdisciplinary programs, PEPS Humain and PEPS MoMIS; the program SYSCOMM of the French National Research Agency, the ANR (project DyXi, grant ANR-08-SYSC-0008) . N.R. was supported by the NSF Grant DMS-1516778.
}

\section*{Author contributions statement}
S.R. collected the raw data and provided the sociological expertise; M.-A.D. and M.B.G. built the database; L.B-G, J.-P.N, H.B. and N.R. were involved in the mathematical modeling; L.B-G performed the numerical analyses and simulations;  L.B-G, J.-P.N, H.B. and S.R. wrote the paper with input from all the authors. All authors reviewed the manuscript.

\small
% \bibliographystyle{ieeetr} 
% \bibliography{riots_biblio}

\clearpage

\newpage

\renewcommand\thefigure{S\arabic{figure}}  
\setcounter{figure}{0} 

\onecolumn

	\noindent {\Large \textbf{Supplementary Information}}\\$\,$\\

\subsection*{Supplementary Figures}

\begin{figure}[hb]
	\begin{minipage}{0.8\linewidth}
		\centering
		\includegraphics[width=0.24\linewidth]{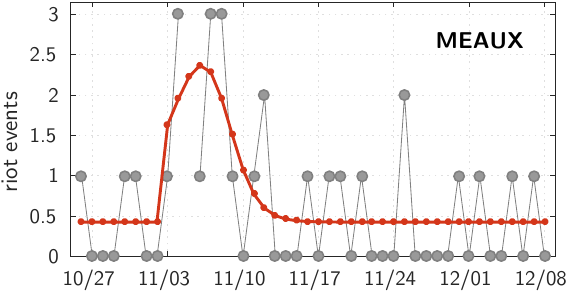} %{img/com_idf/com_77284}
		\includegraphics[width=0.24\linewidth]{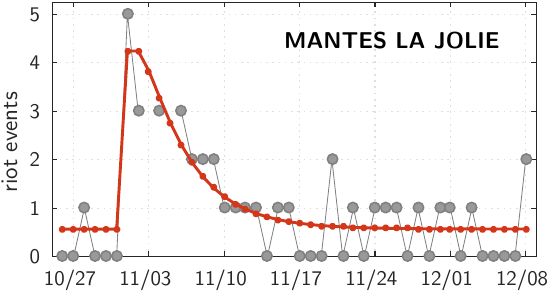}
		\includegraphics[width=0.24\linewidth]{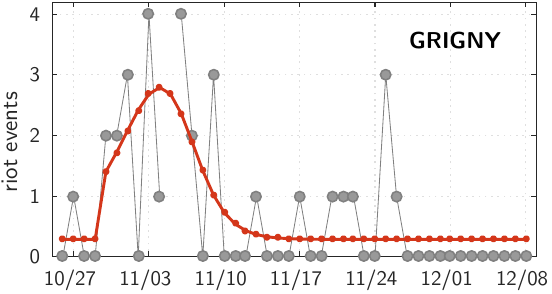}
		\includegraphics[width=0.24\linewidth]{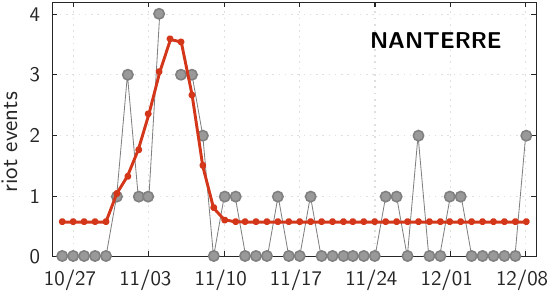}\\
		\vspace{0.2cm}
		\includegraphics[width=0.24\linewidth]{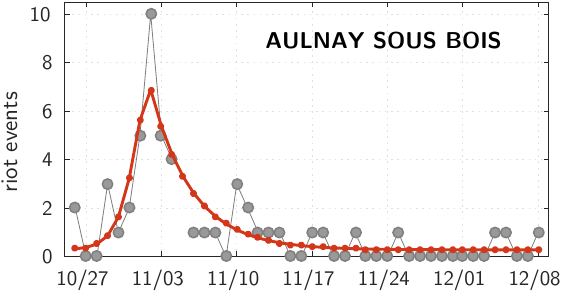}
		\includegraphics[width=0.24\linewidth]{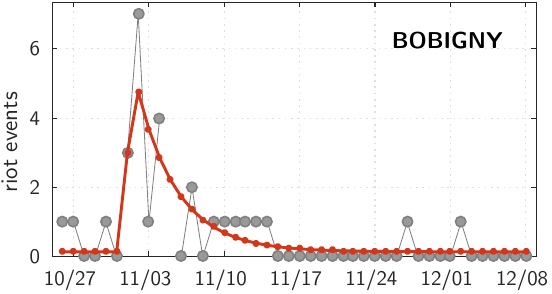}
		\includegraphics[width=0.24\linewidth]{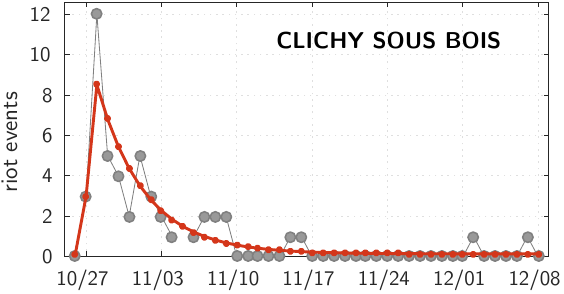}
		\includegraphics[width=0.24\linewidth]{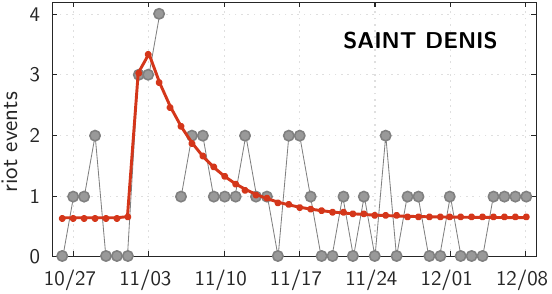}\\
		\vspace{0.2cm}
		\includegraphics[width=0.24\linewidth]{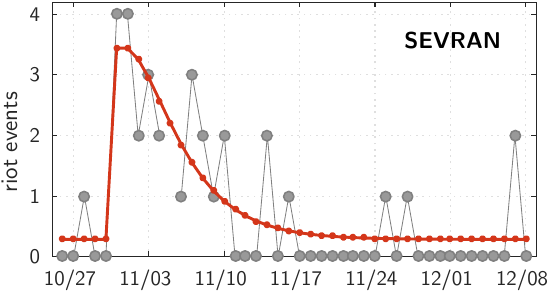}
		\includegraphics[width=0.24\linewidth]{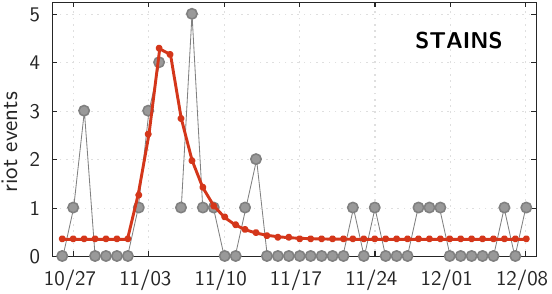}
		\includegraphics[width=0.24\linewidth]{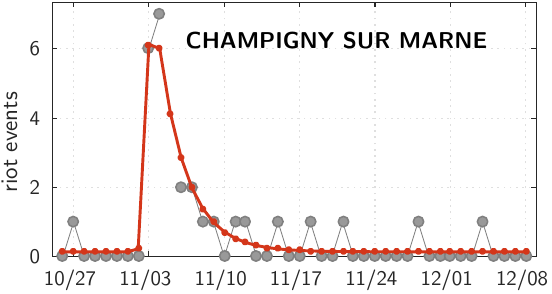}
		\includegraphics[width=0.24\linewidth]{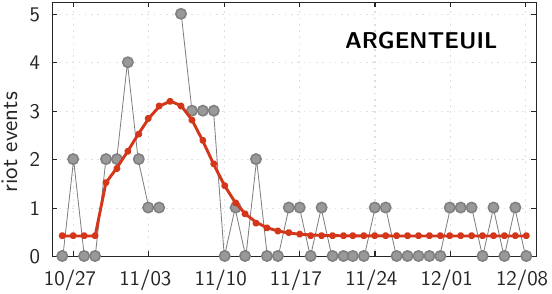}
		\caption{The 2005 French riots: data and single site fits for the 12 most active \^Ile-de-France municipalities. Dots: number of  events. Continuous curve: fit with the single site SIR model.}
		\label{extfig:com_idf_mostactive} % S1
	\end{minipage}
\end{figure}

\begin{figure}
	\begin{minipage}{0.8\linewidth}
		\centering
		\textbf{a}
		\includegraphics[width=0.96\linewidth,valign=t]{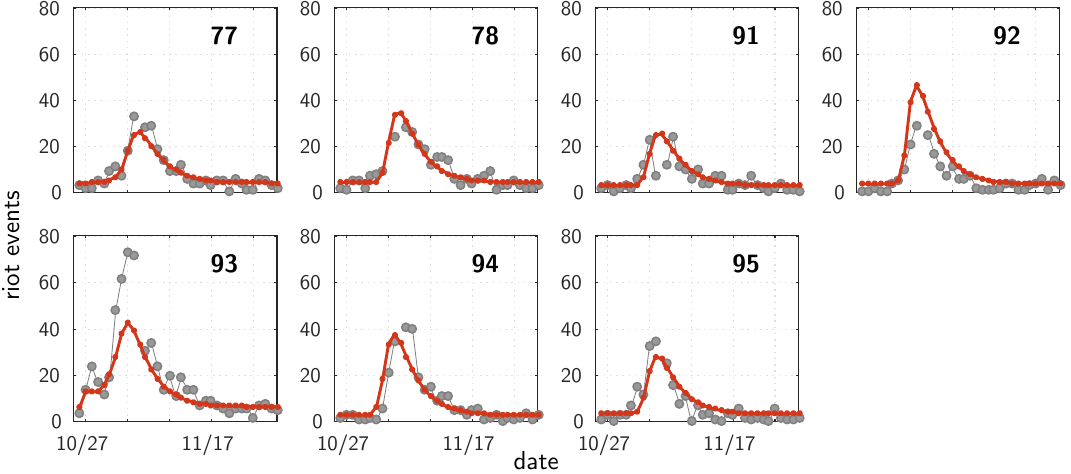}\\
		\vspace{.5cm}
		\textbf{b}
		\includegraphics[width=0.36\linewidth,valign=t]{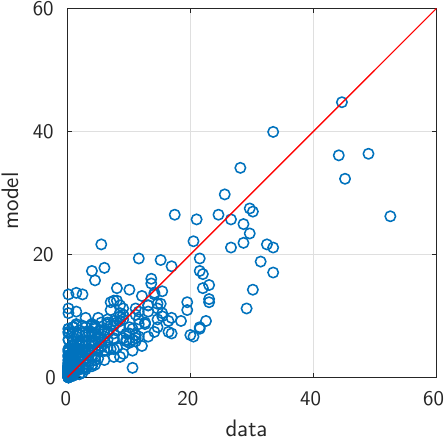}
% 		\includegraphics[width=0.96\linewidth,valign=t]{img/fit_com_idf_bydpt_yabs_control}\\
% 		\vspace{.5cm}
% 		\textbf{b}
% 		\includegraphics[width=0.36\linewidth,valign=t]{img/fit_com_idf_sum_events_control}
		\caption{Control numerical experiment: fit with an inadequate reference population. Here the reference population is taken as the total population.
			(a) \^Ile-de-France municipalities (aggregated by d\'epartements): data (dots), model (continuous curve) making use of the inadequate reference population.
			(b) Total number of events, model vs. data. Each dot represents one municipality. These results should be compared with the one on Fig. 2 of the main text -- apart from the choice of a different reference population, the model options (choice of $\Psi$ and of the weights) are the same.}
		\label{extfig:fit_com_idf_wrongpop}  % S2
	\end{minipage}
\end{figure}

\begin{figure}
	\begin{minipage}{0.8\linewidth}
		\centering
		\includegraphics[width=.98\linewidth]{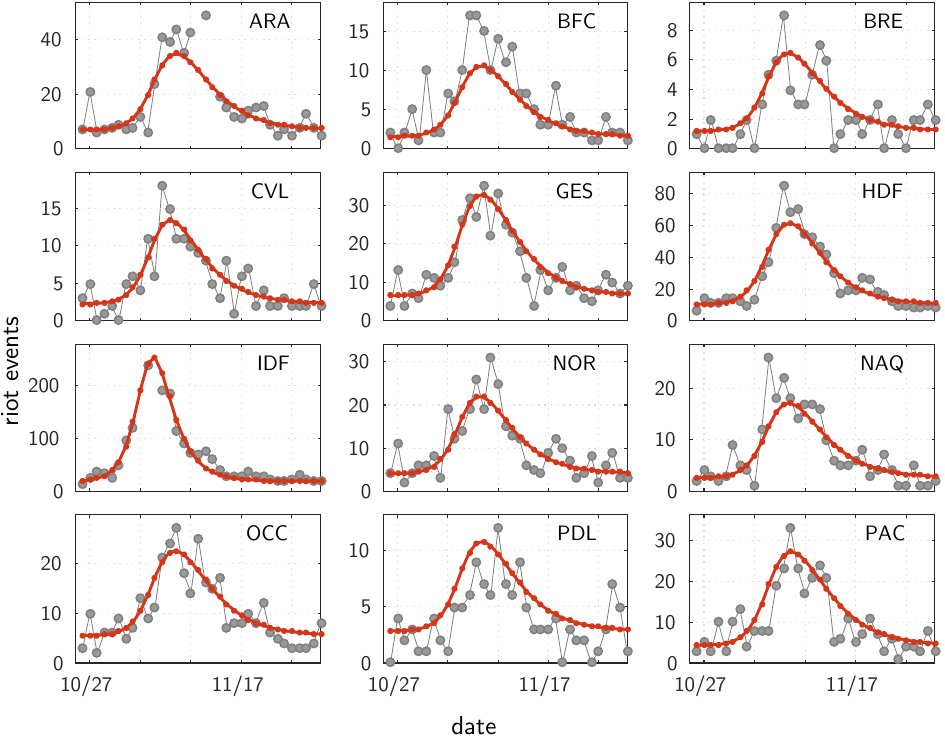} %{img/fit_dpt_reg}
		\caption{Results: all of France, spatial SIR model, calibration at the scale of the d\'epartements. Results aggregated by `régions' (as of 2016), taking into account all the d\'epartements. See main text, section \emph{Fitting the data: the wave across the whole country}. See Supplementary Table S2 for more details on the régions.}
		\label{extfig:linear_nodip_dpt_fr_reg}   % S3
	\end{minipage}
\end{figure}

\begin{figure}
	\begin{minipage}{0.8\linewidth}
		\centering
		\textbf{a}
		\includegraphics[width=0.83\linewidth,valign=t]{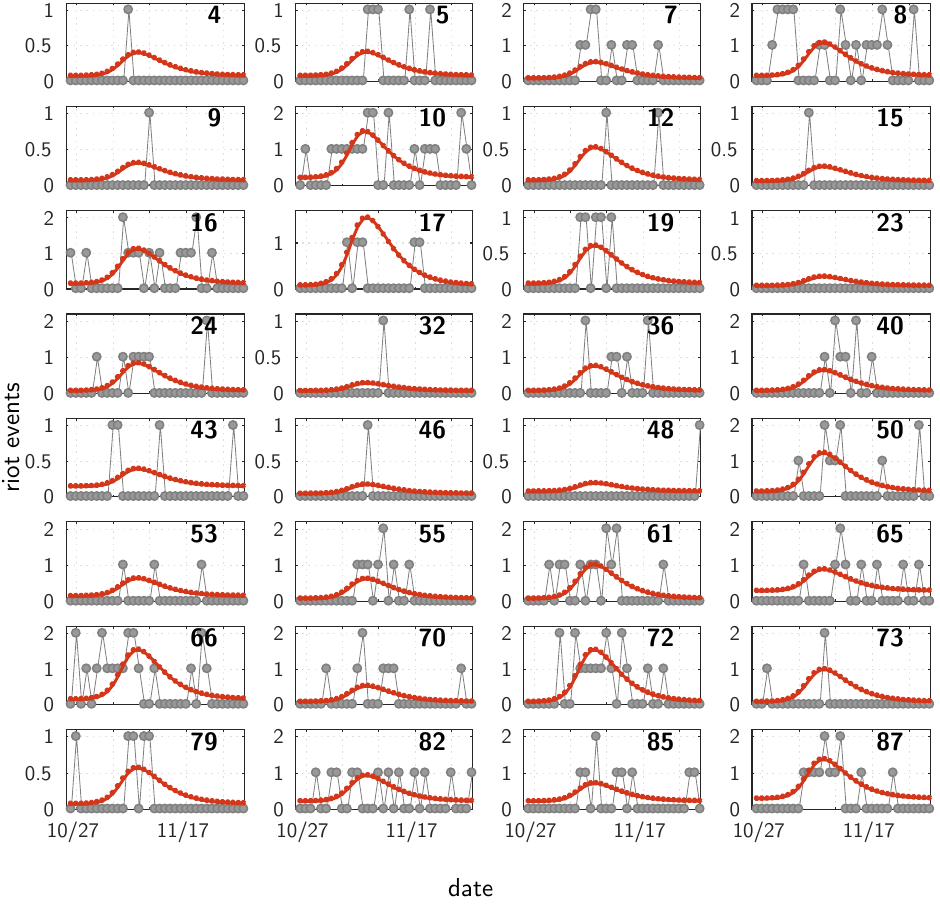}\\ %{img/dpt_fr_minsites}\\
		\vspace{.9cm}
		\textbf{b}
		\includegraphics[width=0.46\linewidth,valign=t]{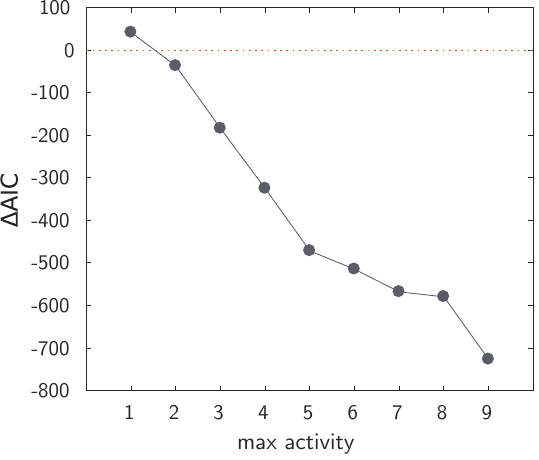} %{img/minorsites_comparisonwithnullmodel}
		\caption{Results, calibration at the scale of the d\'epartements: minor sites. Even where the number of events is very small, the model predicts the sites to be hit by the wave, with a small amplitude and at the correct period of time. 
			%When looking at these figures, one should keep in mind that the model prediction represents the mean value of a stochastic Poisson point process. Thus, for a given day, a value smaller than 1 means that the most probable situation is no event, and we expect either 0 or 1 event.
			(a) All the sites where the number of events on any given day is inferior or equal to 2.
			(b) Comparison between our model and a constant rate null-hypothesis model. The y-axis corresponds to the difference in Akaike Information Criterion. A negative value of $\Delta$AIC means a better performance of our contagion model. The x-axis indicates the maximum activity considered (\textit{ie} all the sites with maximum daily activity inferior or equal to the corresponding value are used in the computation of the difference in AIC). For more details, see Main text, Materials and Methods, section \textit{Minor sites: Comparison with a constant rate null-hypothesis}.
		}
		\label{extfig:dpt_fr_minsites}   % S4
	\end{minipage}
\end{figure}

\begin{figure}
	\begin{minipage}{0.8\linewidth}
		\centering
		\textbf{a}
		\includegraphics[width=0.96\linewidth,valign=t]{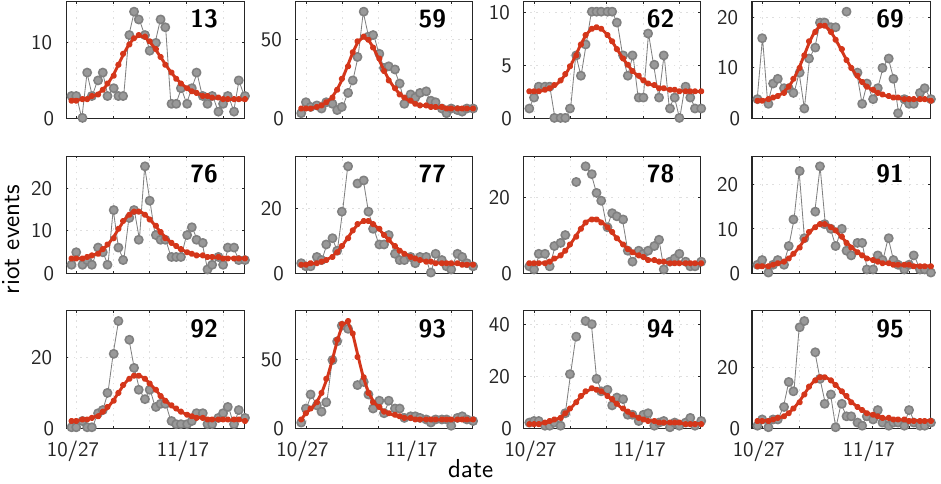}\\ %{img/linear_self_vs_rest_dpt_fr_majsites}\\
		\vspace{.5cm}
		\textbf{b}
		\includegraphics[width=0.36\linewidth,valign=t]{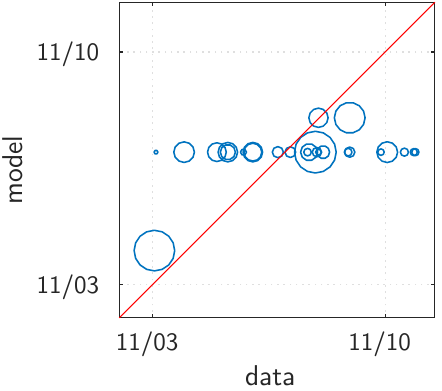} %{img/linear_self_vs_rest_time_max}
		\caption{Distance-independent null hypothesis model. This figure, to be compared with Fig. 4, main text, shows that the absence of geographic dependency in the contagion process fails to reproduce the wave, but also to account for the amplitudes of the riots (see Materials and Methods for details). Plotted here: All of France, model calibration at the scale of the d\'epartements. 
			(a) Time course of the riots in France: data (dots) and model (continuous curves). Only the $12$ most active d\'epartements are shown. 
			(b) Temporal unfolding (date when the number of riot events reaches its maximum value), shown for the d\'epartements having more than $60$ events. Each blue circle has a diameter proportional to the reference population of the corresponding d\'epartements. The red lines depict the identity diagonal line.}
		\label{extfig:fit_dpt_distance_independent}  % S5
	\end{minipage}
\end{figure}

\begin{figure}
	\begin{minipage}{0.8\linewidth}
		\centering
		\textbf{a}
		\includegraphics[width=0.8\linewidth,valign=t]{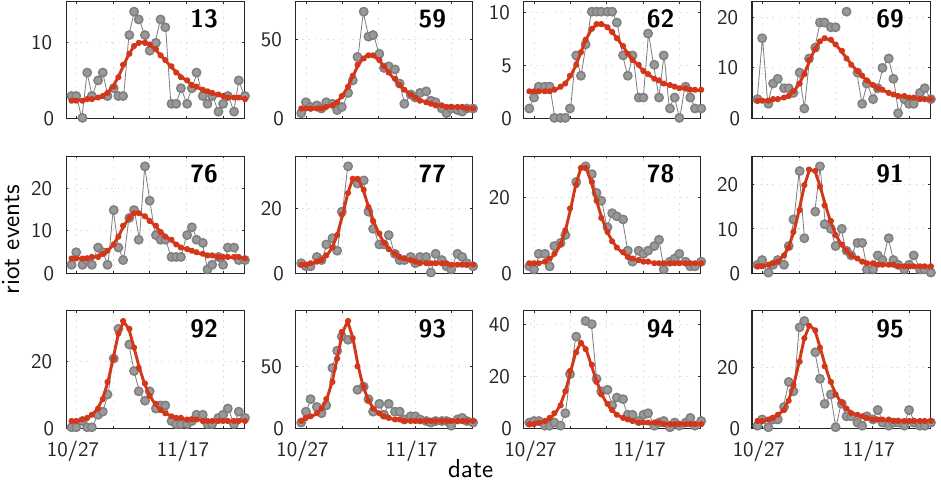}\\ %{img/fit_dpt_majsites}\\
		\vspace{.8cm}
		\textbf{b}
		\includegraphics[width=0.8\linewidth,valign=t]{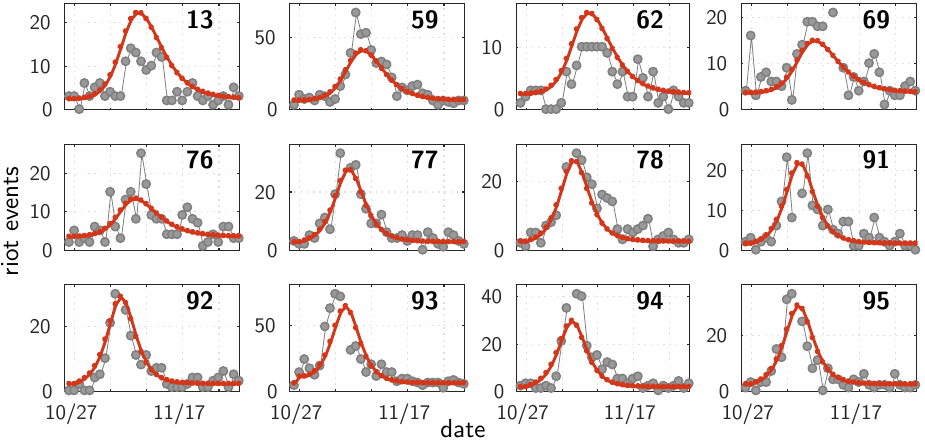}\\ %{img/min_linear_local_global_dpt_fr_majsites}\\
		\vspace{.8cm}
		\textbf{c}
		\includegraphics[width=0.8\linewidth,valign=t]{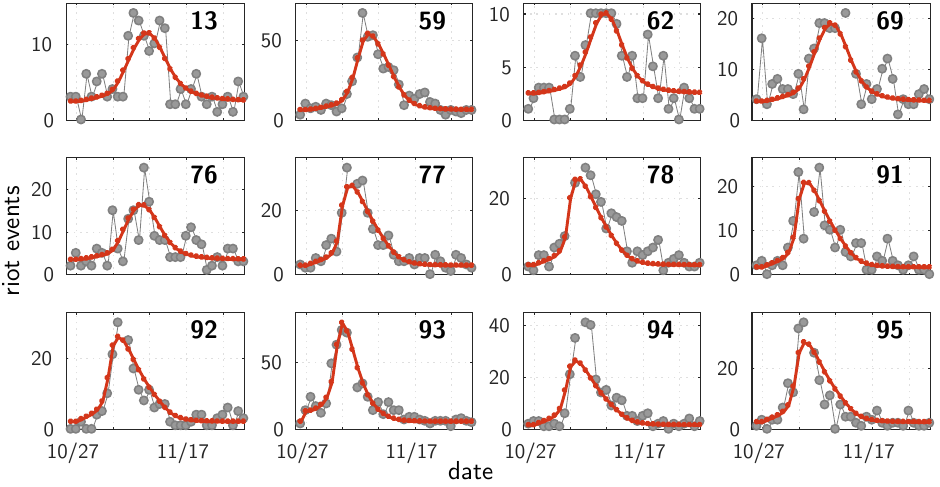} %{img/sigmoid_local_global_dpt_fr_majsites}
		\caption{Model variants. All of France, model calibration at the scale of the d\'epartements. (a) For ease of comparison, reproduction of Fig. 4b (model with 9 free parameters); (b) and (c), same model as (a), except: (b) no extra specific susceptibility value $\beta$  (6 free parameters), (c) the use of a sigmoidal function for $\Psi$ (12 free parameters).}
		\label{extfig:model_variants_dpt}  % S6
	\end{minipage}
\end{figure}

\begin{figure}
	\begin{minipage}{0.8\linewidth}
		\centering
		\includegraphics[width=0.5\linewidth]{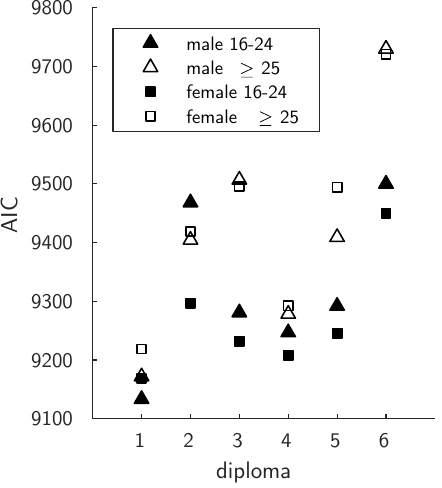} %{img/comparison_dpt_pop}
		\caption{Choice of the reference population.  Model comparison,  considering cross-linked database that involve age, sex and diploma (INSEE statistics of 2006). All of France, model calibration at the scale of the départements. For each point, the model is the minimal one, using the 6 free parameters $\omega$, $A$, $\zeta_0$, $d_0$, $\xi$, $\beta$, and the corresponding reference population. For the Akaike Information Criterion (AIC), the lower is the better. Diploma categories: 1, no diploma; from 2 to 6, higher and higher levels of education, with 5, high school diploma, 6 university diploma (specific French levels: 2: CEP; 3: BEPC; 4: CAP-BEP; 5: Bac général, bac technique; 6: Diplôme universitaire 1er, 2ème ou 3ème cycle, BTS-DUT).}
		\label{extfig:model_comparison_pop}  % S7
	\end{minipage}
\end{figure}

\clearpage	
\subsection*{Supplementary Tables}%$\,$\\

\begin{table}[!h]
	\begin{subtable}{\textwidth}
		\begin{tabular}{p{9.5cm}|p{2.1cm}|p{3.0cm}|c}
			\hline
			Description of the model & \small Number of free parameters & \small List of parameters & AIC\\ 
			\hline\hline
			model presented in the paper (see Fig. 2) \newline 
			$\Psi$ strict threshold, 
			weights $W_{kj}$ power law
			& 8 & $\omega$, $A$, $\zeta_0$, $d_0$, $\delta$, $\eta$, $\gamma$, $\Lambda_c$ & 9718\\
			\hline
			\hline
			same, but with $\Psi$ linear & 6 & $\omega$, $A$, $\zeta_0$, $d_0$, $\delta$, $\beta$  &  9789\\
			\hline
			same, but with weights $W_{kj}$ exponential decay + constant & 8 & $\omega$, $A$, $\zeta_0$, $d_0$, $\xi$, $\eta$, $\gamma$, $\Lambda_c$  & 9738\\
			\hline
			same, but with total population as reference population\newline
			(see Fig.~\ref{extfig:fit_com_idf_wrongpop})& 8 & $\omega$, $A$, $\zeta_0$, $d_0$, $\delta$, $\eta$, $\gamma$, $\Lambda_c$  & 10055\\		
			\hline
		\end{tabular}
		\caption{Model comparison. \^Ile-de-France région, model calibration at the scale of municipalities.}
		\label{exttab:model_comparison_com_idf} 
	\end{subtable}
	
	\vspace{1cm}
	
	\begin{subtable}{\textwidth}
		\begin{tabular}{p{9.5cm}|p{2.1cm}|p{3.0cm}|c}
			\hline
			Description of the model  & \small Number of free parameters & \small List of parameters & AIC\\ 
			\hline\hline
			model presented in the paper (see Fig. 4) \newline 
			$\Psi$ linear,
			weights $W_{kj}$ exponential decay + constant\newline
			outliers defined with a three standard deviations criterion
			& 9 & $\omega$, $A$, $\zeta_0$, $d_0$, $\xi$, $\beta$, $\beta_{13}$, $\beta_{62}$, $\beta_{93}$ & 8979\\
			\hline
			\hline
			same, but weights $W_{kj}$ power law & 9 &  $\omega$, $A$, $\zeta_0$, $d_0$, $\delta$, $\beta$, $\beta_{13}$, $\beta_{62}$, $\beta_{93}$ &  9000\\
			\hline
			same, but weights $W_{kj}$ self vs rest \newline
			(distance-independent null hypothesis model see Fig.~\ref{extfig:fit_dpt_distance_independent}) & 8 &  $\omega$, $A$, $\zeta_0$, $\xi$, $\beta$, $\beta_{13}$, $\beta_{62}$, $\beta_{93}$ &  9419\\
			\hline
			same, but with densities instead of numbers & 9 &  $\omega$, $A$, $\zeta_0$, $d_0$, $\xi$, $\beta$, $\beta_{13}$, $\beta_{62}$, $\beta_{93}$ &  9080\\
			\hline
			same, but no extra susceptibility value $\beta$
			(see Fig.~\ref{extfig:model_variants_dpt}b) & 6 &  $\omega$, $A$, $\zeta_0$, $d_0$, $\xi$, $\beta$ &  9133\\
			\hline
			same, but with one susceptibility value $\beta$ \newline
			outliers defined with a four standard deviations criterion
			& 7 &  $\omega$, $A$, $\zeta_0$, $d_0$, $\xi$, $\beta$, $\beta_{13}$ &  9058\\
			\hline
			same, but with $\Psi$ sigmoidal\newline
			(see Fig.~\ref{extfig:model_variants_dpt}c) & 12 &  $\omega$, $A$, $\zeta_0$, $d_0$, $\xi$, $\beta$, $\tau$, $\gamma$, $\Lambda_c$, $\beta_{13}$, $\beta_{62}$, $\beta_{93}$ & 8923\\
			\hline
		\end{tabular}
		\caption{Model comparison. All of France, model calibration at the scale of the départements.}
		\label{exttab:model_comparison_dpt} 
	\end{subtable}
	\caption{Summary of miscellaneous model variants that were tested. In both sub-tables, ``same'' means same model as the one presented in the main text and described in the first row, except for the described option. 
		See Material and Methods for more details. For convenience, we give below a brief reminder of the equations of some options for the interaction  weights ($W_{kj}$ in $\Lambda_k(t) = \sum_j W_{kj} \lambda_j(t)$), and for the function $\Psi$ used to model the probability for a susceptible agent to become a rioter.}
	\begin{minipage}{\linewidth}
		%\small
		Function $\Psi$:
		\begin{itemize}
			\item linear: $\Psi(\Lambda) =  \beta \Lambda$
			\item strict threshold: $\Psi(\Lambda) = 0$ if $\Lambda \leq \Lambda_{c}$; $\Psi(\Lambda) = \eta \left(1-\exp{-\gamma\;(\Lambda-\Lambda_{c})}\right)$ if $\Lambda > \Lambda_{c}$
			\item sigmoidal: $\Psi(\Lambda) = \beta (1-\exp{-\tau\Lambda})\left(1-\exp{-\gamma\;(\Lambda-\Lambda_{c})}\right)^{-1}$
		\end{itemize}		
		Weights: 
		\begin{itemize}
			\item power law: $W_{kj} = \left(1+\text{dist}(k,j)/d_0\right)^{-\delta}$
			\item exponential decay + constant: $W_{kj} = \xi\;+\; (1-\xi) \exp\left(-\text{dist}(k,j)/d_0\right)$
			\item self vs rest: $W_{kj} = 1$ if $j=k$; $W_{kj} =\xi$ otherwise.
		\end{itemize}		
	\end{minipage}
\end{table}

\begin{table}[hb]
	\begin{tabular}{l|l|l}
		code & full name & départements\\
		\hline
		ARA & Auvergne-Rhône-Alpes & 01 03 07 15 26 38 42 43 63 69 73 74\\
		BFC & Bourgogne-Franche-Comté & 21 25 39 58 70 71 89 90\\
		BRE & Bretagne & 22 29 35 56\\
		CVL & Centre-Val de Loire & 18 28 36 37 41 45\\
		GES & Grand Est & 08 10 51 52 54 55 57 67 68 88\\
		HDF & Hauts-de-France & 02 59 60 62 80\\
		IDF & Île-de-France & 75 77 78 91 92 93 94 95\\
		NOR & Normandie & 14 27 50 61 76\\
		NAQ & Nouvelle-Aquitaine & 16 17 19 23 24 33 40 47 64 79 86 87\\
		OCC & Occitanie & 09 11 12 30 31 32 34 46 48 65 66 81 82\\
		PDL & Pays de la Loire & 44 49 53 72 85\\
		PAC & Provence-Alpes-Côte d'Azur & 04 05 06 13 83 84\\
	\end{tabular}
	\caption{Code ISO 3166-2 (without the prefix FR-) and associated name of the 12 régions of Metropolitan France (as of 2016, excluding Corsica), along with the ID number of the départements that they encompass.}
	For the ISO codes, see \url{https://www.iso.org/obp/ui/#iso:code:3166:FR}.
	\label{exttab:region_names}  % S2
\end{table}

\clearpage	
\twocolumn
\subsection*{Supplementary Videos}%$\,$\\
\vspace{0.0cm}
\noindent $\bullet$ Supplementary Video 1.
{\bf Riot propagation around Paris: smoothed data.}\\
This video shows the riot propagation around Paris. The map shows the municipality boundaries, with Paris at the center. For each municipality for which data is available, a circle is drawn with an area proportional to the estimate of the size of the susceptible population (see main text, section Methods). Instead of making use of the raw data, for each municipality we replaced each day value by the one given by  the fit with the single site epidemic model considered here as a tool for smoothing the data. The color represents the intensity of the rioting activity: the warmer the color, the higher the activity. The pace of the video corresponds to three days per second. In order to improve the fluidity of the video, we increased the number of frames per second by interpolating each day with 7 new frames, whose values are computed thanks to a piece-wise cubic interpolation of the original ones. The resulting frame rate is then 24 frames per second.\\
The maps have been generated with the Mapping toolbox of the MATLAB software making use of the Open Street Map data \copyright OpenStreetMap contributors (\url{https://www.openstreetmap.org/copyright}).\\
The video is encoded with the open standard H.264.\\

\vspace{-0.1cm}
\noindent $\bullet$ Supplementary Video 2. 
{\bf Riot propagation around Paris: model with  non-local contagion.}\\
This video shows the riot activity as predicted by the data-driven global epidemic-like model. Same technical details as for the SI Video 1.\\
The maps have been generated with the Mapping toolbox of the MATLAB software making use of the Open Street Map data \copyright OpenStreetMap contributors (\url{https://www.openstreetmap.org/copyright}).\\
The video is encoded with the open standard H.264.\\

\vspace{-0.1cm}
\noindent $\bullet$ Supplementary Video 3.
{\bf Spatial SIR: wave propagation in a homogeneous medium.}\\
We illustrate the formal continuous spatial SIR model with a video showing the propagation of a wave. The underlying medium is characterized by a uniform density of susceptible individuals. The weights $w(x-y)$ in the interaction term are given by a decreasing exponential function of the Euclidean distance $||x-y||$.\\
The video is encoded with the open standard H.264.\\

\vspace{-0.1cm}
\noindent $\bullet$ Supplementary Video 4. 
{\bf Spatial SIR: wave propagation in a non-homogeneous medium.}\\
Same as SI Video 3, but with a heterogeneous density of susceptible individuals, characterized by (1) a decrease of the density towards 0 near the boundary of the image, so that the wave dies before leaving the frame, (2) a hole at the center, which is then bypassed by the wave, (3) a concentration of susceptible individuals that globally decreases on the y-axis, so that the wave dies while going downward.\\
The video is encoded with the open standard H.264.

\begin{figure}
	\centering
	\vspace{3.1cm}
	\includegraphics[width=.9\linewidth]{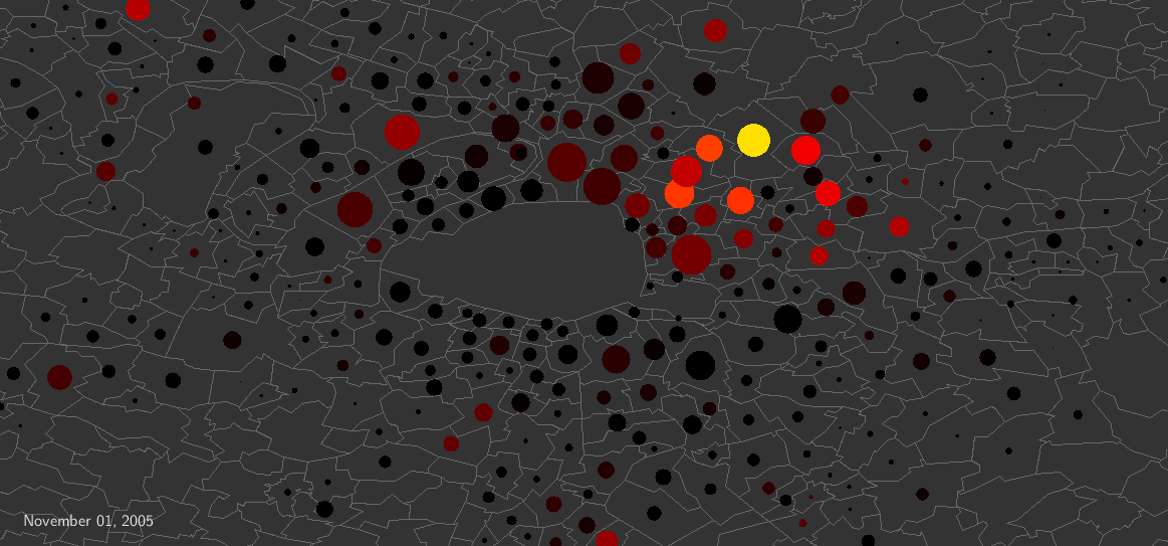}\\
	{\footnotesize Still image from SI Video 1}\\
	\vspace{2.8cm}
	\includegraphics[width=.9\linewidth]{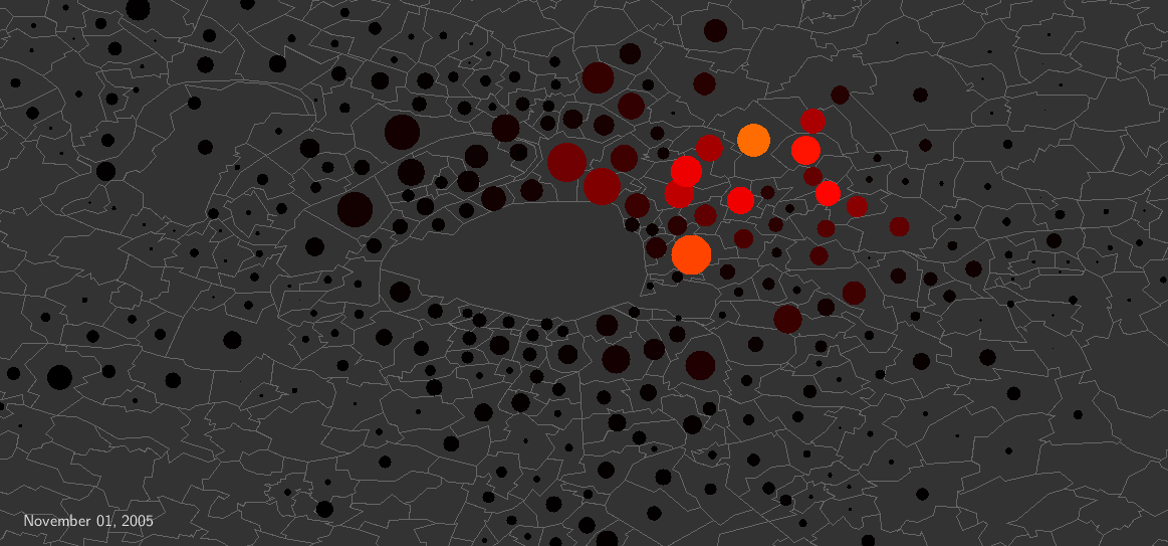}\\
	{\footnotesize Still image from SI Video 2}\\ 
	\vspace{0.7cm}
	\includegraphics[width=.3\linewidth]{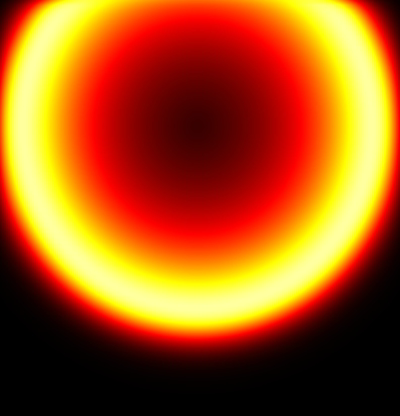}\\
	{\footnotesize Still image from SI Video 3}\\
	\vspace{0.7cm}
	\includegraphics[width=.3\linewidth]{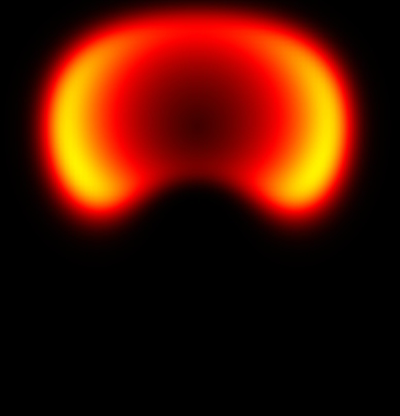}\\
	{\footnotesize Still image from SI Video 4}
\end{figure}

\end{document}